\documentclass[10pt,a4paper,twoside]{article}
\usepackage[]{natbib}
\usepackage{hyperref}
\usepackage{graphicx}
\usepackage{helvet}
\usepackage{times} 
\usepackage{afterpage}
\usepackage{comment}
\usepackage{textcomp}
\usepackage{fancyhdr} 
\usepackage{lastpage}
\usepackage{floatrow}
\usepackage[tableposition=top]{caption}
\usepackage{hyperref}
\hypersetup{
    bookmarks=true,         
    unicode=false,          
    pdftoolbar=true,        
    pdfmenubar=true,        
    pdffitwindow=false,     
    pdftitle={Certificate},    
    pdfauthor={Dr. Dimitrios Gouliermis},     
    pdfsubject={Astronomy},   
    pdfcreator={Dr. Dimitrios Gouliermis},   
    pdfproducer={},  
    pdfnewwindow=true,      
    colorlinks=false,       
    linkcolor=blue,          
    citecolor=green,        
    filecolor=magenta,      
    urlcolor=cyan           
}

\usepackage{float}
\floatstyle{plaintop}
\restylefloat{table}

\usepackage{multicol}

\makeatletter
\renewenvironment{thebibliography}[1]
     {\begin{multicols}{2}[\section*{\refname}]%
      \@mkboth{\MakeUppercase\refname}{\MakeUppercase\refname}%
      \list{\@biblabel{\@arabic\c@enumiv}}%
           {\settowidth\labelwidth{\@biblabel{#1}}%
     \scriptsize
      \itemindent=-.5em
      \itemsep=-5pt 
                  \leftmargin\labelwidth
            \advance\leftmargin\labelsep
            \@openbib@code
            \usecounter{enumiv}%
            \let\p@enumiv\@empty
            \renewcommand\theenumiv{\@arabic\c@enumiv}}%
      \sloppy
      \clubpenalty4000
      \@clubpenalty \clubpenalty
      \widowpenalty4000%
      \sfcode`\.\@m}
     {\def\@noitemerr
       {\@latex@warning{Empty `thebibliography' environment}}%
      \endlist\end{multicols}}
\makeatother

\long\def\symbolfootnote[#1]#2{\begingroup%
\def\thefootnote{\fnsymbol{footnote}}\footnote[#1]{#2}\endgroup}

\usepackage{booktabs,caption,fixltx2e}
\usepackage[flushleft]{threeparttable}

\newcommand{\beq}{\begin{equation}}
\newcommand{\eeq}{\end{equation}}

\long\def\symbolfootnote[#1]#2{\begingroup%
\def\thefootnote{\fnsymbol{footnote}}\footnote[#1]{#2}\endgroup}

\def\solar{\mbox{$_{\normalsize\odot}$}}

\def\micron{\hbox{$\mu {\rm m}$}}

\newcommand{\AmS}{{\protect\the\textfont2
  A\kern-.1667em\lower.5ex\hbox{M}\kern-.125emS}}
\newcommand{\lsim}{\ \raise
-2.truept\hbox{\rlap{\hbox{$\sim$}}\raise5.truept\hbox{$<$}\ }}
\newcommand{\gsim}{\ \raise
-2.truept\hbox{\rlap{\hbox{$\sim$}}\raise5.truept\hbox{$>$}\ }}
\newcommand{\simsim}{\ \raise
-2.truept\hbox{\rlap{\hbox{$\sim$}}\raise5.truept\hbox{$\sim$}\ }}

\newcommand{\itemhscd}{\noindent\hspace*{.75cm}\raise-2.5truept\hbox{\Huge$\cdot$}\ }

       \usepackage{enumitem,amssymb}
       \newlist{todolist}{itemize}{2}
       \setlist[todolist]{label=$\square$}
       \usepackage{pifont}
       \usepackage{titlesec}
       \usepackage[labelsep=endash]{caption}

       \titleformat{\section}{\sffamily\large\bfseries}{\thesection.}{1em}{\sffamily\large\bfseries}
       \titleformat{\subsection}{\sffamily\normalsize\bfseries}{\thesubsection}{1em}{\sffamily\normalsize\bfseries}
       \titleformat{\subsubsection}{\sffamily\small\bfseries}{\thesubsubsection}{1em}{\sffamily\small\bfseries}
       \titleformat{\paragraph}{\sffamily\footnotesize\bfseries}{\theparagraph}{1em}{\sffamily\footnotesize\bfseries}\titlespacing*{\paragraph}{0pt}{3.25ex plus 1ex minus .2ex}{1.5ex plus .2ex}

\usepackage{color}
\usepackage{soul}
    \definecolor{lightgray}{gray}{0.85}
    \sethlcolor{lightgray}
    
\long\def\greybox#1{%
    \newbox\contentbox%
    \newbox\bkgdbox%
    \setbox\contentbox\hbox to \hsize{%
        \vtop{
            \kern\columnsep
            \hbox to \hsize{%
                \kern\columnsep%
                \advance\hsize by -2\columnsep%
                \setlength{\textwidth}{\hsize}%
                \vbox{
                    \parskip=\baselineskip
                    \parindent=0bp
                    #1
                }%
                \kern\columnsep%
            }%
            \kern\columnsep%
        }%
    }%
    \setbox\bkgdbox\vbox{
        \pdfliteral{0.85 0.85 0.85 rg}
        \hrule width  \wd\contentbox %
               height \ht\contentbox %
               depth  \dp\contentbox
        \pdfliteral{0 0 0 rg}
    }%
    \wd\bkgdbox=0bp%
    \vbox{\hbox to \hsize{\box\bkgdbox\box\contentbox}}%
    \vskip\baselineskip%
}

\def\aj{AJ}                   
\def\araa{ARA\&A}             
\def\apj{ApJ}                 
\def\apjl{ApJL}                
\def\apjs{ApJS}               
           
\def\apss{Ap\&SS}             
\def\aap{A\&A}                
          
\def\aaps{A\&AS}

\def\mnras{MNRAS}

\def\pasp{PASP}               
\def\pasj{PASJ}

\def\sovast{Soviet~Ast.}

\def\pasa{PASA}

\def\nar{NewAR}

\begin{document}

\rmfamily

\setlength{\baselineskip}{0.605cm}

\noindent {\Large\bfseries Unbound Young Stellar Systems: Star Formation on the loose}\\
\vspace*{-.35truecm}\\

\noindent{\large\bfseries Dimitrios A. Gouliermis}\\
\vspace*{-.35truecm}\\
\noindent{\bfseries Zentrum f\"ur Astronomie der Universit\"at Heidelberg, Institut f\"ur Theoretische Astrophysik,}\\ 
\noindent{\bfseries Albert-Ueberle-Str.\,2, 69120 Heidelberg, Germany}\\
\vspace*{-.35truecm}\\
\noindent{\href{mailto:gouliermis@uni-heidelberg.de}{gouliermis@uni-heidelberg.de}}\\
\hrule
 

\begin{abstract}
\noindent 
Unbound young stellar systems, the loose ensembles of physically related young bright stars, {trace the typical regions of recent star formation in galaxies.} Their morphologies vary from small few pc-size  associations of newly-formed stars to enormous few kpc-size complexes composed of stars few 100 Myr old. These stellar conglomerations are located within the disks and along the spiral arms and rings of star-forming disk galaxies, and they are the active star-forming centers of dwarf and starburst galaxies. Being associated with star-forming regions of various sizes, these stellar structures {trace the regions where stars form} at various length- and time-scales, from compact clusters to whole galactic disks. Stellar associations, the prototypical unbound young systems, and their larger counterparts, stellar aggregates, and stellar complexes, have been the focus of several studies for quite a few decades, with special interest on their demographics, classification, and structural morphology. The compiled surveys of these loose young stellar systems demonstrate that the clear distinction of these systems into well-defined classes is not as straightforward as for stellar clusters, due to their low densities, asymmetric shapes and variety in structural parameters. These surveys also illustrate that unbound stellar structures follow a clear hierarchical pattern in the clustering of their stars across various scales. Stellar associations are characterized by significant sub-structure with bound stellar clusters being their most compact parts, while associations themselves are the brighter denser parts of larger stellar aggregates and stellar complexes, which are members of larger super-structures up to the scale of a whole star-forming galaxy. This structural pattern, which is usually characterized as self-similar or fractal, appears to be identical to that of star-forming giant molecular clouds and interstellar gas, driven mainly by turbulence cascade. In this short review, I make a concise compilation of our understanding of unbound young stellar systems across various environments in the local universe, as it is developed during the last 60 years. I present a factual assessment of the clustering behavior of star formation, as revealed from the assembling pattern of stars across loose stellar structures and its relation to the interstellar medium and the environmental conditions. I also provide a consistent account of the processes that possibly play important role in the formation of unbound stellar systems, compiled from both theoretical and observational investigations on the field.


\end{abstract}
\vspace*{.25truecm}
\hrule
\vspace*{.25truecm}



\tableofcontents

\section{Introduction\label{s:intro}}

{\sl The vast majority of stars form in clusters}. This concept has emerged from observations in the Milky Way, showing that ``embedded clusters account for the 70\,--\,90\% fraction of all stars formed in Giant Molecular Clouds (GMCs)" \citep{Lada2003ARAA}. These findings led to the generalization of the star formation process as being typically a {\sl clustered} process, and loose OB associations were considered to be primarily the remnants of dissolved embedded clusters \citep{Lada2003ARAA}. 
However, galaxies of the Local Group, including our own Milky Way, show over-densities\footnote{In the context of this review, the term `over-density' of stars is used to describe stellar concentrations with densities (surface or volume) well above the average values of that in their surrounding field.} of early-type stars, which are large, extending at scales from few 10 to several 100 parsecs, and loose in appearance. 
In fact, these structures host in their vicinities the bulk of young massive stars in a galaxy \citep[only \lsim\,10\% of newly-formed stars are confined in gravitationally bound clusters; e.g.,][]{Schweizer2009, BonattoBica2011c, FallChandar2012}. They are located within the thin disks and along the spiral arms and rings of star-forming disk galaxies, and they are the complex star-forming centers of irregular and star-burst galaxies. Their morphology implies that these stellar systems are gravitationally unbound, i.e., have a higher (more positive) gravitational potential energy than the sum of their parts, which makes them dynamically unstable. As a consequence, the fact that they host large numbers of distributed massive blue stars implies either that stars can also form -- at least partially -- in a {\sl dispersed} (unbound) fashion, or that indeed {\sl all} stars formed in clusters, but are immediately dislocated from their compact birthplaces through evaporation of their unstable systems. Extended loose stellar systems, being signposts of large-scale star formation, naturally attract our interest in understanding the formation of stars across galaxies. This in turn allows the addressing of important issues, such as the stellar Initial Mass Function (IMF) and its variability, the role of feedback in pausing star formation or triggering subsequent generations of stars, and the dynamics of star-forming molecular clouds producing both bound and unbound systems. 


In this article I shortly review the current state of empirical and theoretical knowledge concerning loose concentrations of young stars, and explore our current understanding of processes that influence the formation and evolution of these stellar structures. The term {\sl ``young stellar groupings''} is coined by \cite{Efremov1989} to describe the whole range of young stellar systems between small compact clusters and large loose structures. Throughout this article, I use the term {\sl Unbound Young Stellar Systems} (UYSS) to describe only gravitationally unbound stellar conglomerates, due to their notable structural differences from typical bound star clusters. This class of stellar concentrations includes a quite large range of objects, which extend at various length scales and at various degrees of self binding; from small (semi-)compact loose systems named {\sl associations}, to huge extended superstructures of massive young stars that make up whole parts of galactic spiral arms, known as {\sl stellar complexes}\footnote{Stellar complexes are referred to in the literature by various other terms, such as cluster, star-forming, CO, cloud, and star complexes.}. What is very interesting about UYSS of different length scales is that they are not independent from each other. Being the nurseries of star cluster formation, they are not disconnected from compact clusters either. It is well documented that compact clusters occupy the high-density peaks of stellar associations and aggregates\footnote{The class of ``stellar aggregates'' as an intermediate type of groupings with diameters of 200\,-\,300\,pc was introduced also by \cite{Efremov1989} to bridge the gap between the 1-kpc scale complexes and the 100-pc scale associations.}
The latter systems are basically the ``more compact parts'' of large stellar structures \citep[e.g.,][]{Elmegreen2006HierarchicalSF}. Few such structures can further build up {\sl superstructures} of stars and gas, i.e., the so-called {\sl super-complexes} of galactic arms \citep[e.g.,][]{Efremov2015A&AT}. This stellar clustering pattern, with large loose stellar constellations comprising smaller more compact (but still unbound) stellar ensembles, which themselves ``break-up'' to even smaller and compact structures, is known as {\sl hierarchical} \citep{Larson1994, ElmegreenEfremov1996}. 

This stellar clustering behavior implies that young stellar groupings of different types are naturally linked to each other across a wide range of physical scales, and indicates an underlying hierarchical pattern in the star formation process. Under these circumstances, the classification of stellar groupings into clusters, associations, aggregates, may appear `heuristic', since it based on morphological arguments, but it is also practical, because it reflects the variety of stellar systems in terms of self-gravity as a natural effect of {\sl hierarchical star formation} \citep{Elmegreen2009ApSS}. While this distinction may be very intuitive for the separation of bound (i.e., star clusters) from unbound (i.e., stellar associations) systems, it is not entirely clear what is the balance between localized star formation and global galactic dynamics in influencing the formation and evolution of larger aggregates or even complexes. Therefore, the origin of the hierarchical clustering pattern, and its relation to the formation and survival of loose stellar systems, will be also addressed in this article.

The present review focuses mainly on stellar associations and extended stellar groupings in various galaxies of the Local Group, including the Milky Way. For dedicated reviews on Galactic unbound associations I direct the reader to the NATO ASI paper by \cite{Brown1999ASIC}, which is very complete concerning OB associations located within $\sim$\,1.5\,kpc from the Sun, and the review article by \cite{Blaauw1964}, which is also still an essential reading. I also refer the reader to the {\sl Handbook of Star Forming Regions} \citep{Reipurth2008hsf1, Reipurth2008hsf2}, where a wealth of information is provided about giant star-forming regions (or star-forming complexes) in the Milky Way. Recommended review articles concerning OB associations in galaxies of the Local Group are those focusing on two topics that will not be sufficiently addressed here, namely their massive stellar content \citep{Garmany1994, Massey2013} and the corresponding IMF \citep{Massey98a}. 

The article is structured so that to include  in Section\,\ref{s:define} a historical overview of studies of UYSS and to address their definition. The most complete account to date of surveys of such systems is also made in the same section, where the question of the universality of the characteristic length-scale for star formation, is also addressed with the use of data from these surveys. In Section\,\ref{s:hierarchy} the hierarchical clustering pattern of UYSS is discussed in terms of the scale-free nature of star formation and the fractal structure of the interstellar matter. The clustering behavior of UYSS across various scales and galactic environments is also addressed in this section. Section\,\ref{s:usf} focuses on the origins of UYSS, based on evidence from  observational and theoretical studies of unbound star formation and its connection to clustered star formation. Finally, a short summery with concluding remarks is given in Section\,\ref{s:conclusion}.


\section{Historical Overview \label{s:define}}

Extended conglomerates of young blue stars are the signposts of the most recent star formation across galactic spirals and bursty irregular galaxies in the local universe. Their prominent appearance in blue photometric plates since the middle of the last century motivated the study of these stellar systems in order to understand their connections to the galactic structure and the star formation process. {Atlases of stars, clusters and interstellar objects in the Large and Small Magellanic Clouds \citep{HodgeLMCAtlas, HodgeSMCAtlas}, the Andromeda galaxy \citep{HodgeAndromedaAtlas}, and other galaxies of the Local Group \citep{Hodge2002Atlas} demonstrate these outstanding stellar constellations.} However, the systematic identification of loose stellar structures and their distinction from their surroundings is not trivial. 
In the Milky Way, where the young stellar populations of the disk suffer from field contamination due to the depth along the line-of-site, kinematic information is crucial in identifying individual loose structures of blue stars as co-moving OB associations \citep[][]{deZeeuw1999}. On the other hand, it is relatively straightforward to visually distinguish over-densities of young bright stars from their galactic stellar environments in other galaxies, although establishing the identity of these structures presents its own issues \citep{Hodge1986, Kontizas1994}. As a consequence, several investigations are focused on the compilation of surveys of loose young stellar groupings in galaxies of the Local Group, but these surveys show that the construction of a concrete identification scheme for UYSS is rather controversial due to inconsistencies between classification criteria, observational data and applied techniques. Sometimes, the distinction of systems in different classes is not even physically meaningful, since members of one class are not independent from those of another. Nevertheless, the construction of surveys of loose stellar systems in various environments, based upon various criteria, has a significant historical value. It is from these surveys, where the realization of the hierarchical clustering of young stars (groupings within groupings) originates \citep[e.g.,][]{Battinelli1996}. This section presents a historical overview on the basic definitions of loose young systems in the local universe. Their identification through various techniques and with various datasets based on this definition is also discussed, and the concept of a scale-free pattern in their formation is introduced, which in fact makes the distinction of such objects into a well-separated set of classes questionable.


\subsection{Definitions}

Victor Ambartsumian in his lectures delivered at University College London in 1954, presented results on ``a new type of stellar systems'', which are characterized by their instability and rapid expansion in space \citep{Ambartsumian1955}. The spatial extend of these systems is explained by their ``total kinetic energy being much larger than the absolute value of the energy of gravitational interaction''.  In these lectures Ambartumian sets the stage for the investigation of {\sl stellar systems of positive total energy}. The concept of {\sl stellar associations} as the main representatives of such systems was originally introduced by \cite{Ambartsumian1947}, who later separates them into OB- and T-associations \citep{Ambartsumian1968}, based on their dominant stellar content, and relates them to star-forming regions. 


Typically, associations are unbound groups of OB stars, which are younger and more diffuse than open clusters. Their stellar mass volume densities are of the order of \lsim\,0.1\,M{\solar}\,pc$^{-3}$ \citep{Blaauw1964} and their dimensions range from those of ordinary Galactic clusters to a few hundred parsecs \citep{Garmany1994, Lada2003ARAA}. 
They were traditionally identified as extended clusterings of bright OB-type stars with evidence of a common origin, related to HII regions \citep{Morganetal1952}. The common origin of the stellar content of OB associations added a significant constrain to their definition, since 
it postulates that their  stars  should be {\sl coeval}, i.e., they must be born from the same star formation event \citep{Kontizas1999IAU}. However, as discussed latter, recent observations of Galactic associations indicate that they may include more than one stellar generations. Since OB associations are very young, with ages as young as $\sim 1$\,Myr, we can also assume that their unbound nature is primordial, i.e., by birth. However, as discussed in Sect.\,\ref{s:usf}, the formation of unbound associations may not be entirely unrelated to the formation of compact stellar systems.

In some cases, OB associations show considerable {\sl sub-structure}, encompassing various `OB subgoups' \citep{Blaauw1964}. The expansion of these subgroups about the center of the star-forming region suggests that they are unbound from one another \citep{Blaauw1952}. Based on these characteristics {\sl a stellar association is considered to be a sub-structured stellar concentration, produced by star formation events as they  take place along a star-forming cloud}. Recent evidence that supports this picture is discussed in Sect.\,\ref{s:hierarchy}. Star formation across such a structure is displayed by the various related HII regions, formed due to the ionization of the first early-type stars, and being spread along the parental molecular cloud. This, however, suggests that these systems {\sl may not be necessarily coeval}, but they can exhibit a spread of ages as large as 10 Myr \citep[e.g.,][]{Soderblom2014ppvi}.  Young loose systems with dimensions larger than the `typical' stellar associations, comprising a variety of sub-systems (from one to few Myr old) are proposed as a separate class of loose systems named {\sl stellar aggregates} \citep[or OB-aggregates;][]{Efremov1989}. Aggregates are usually connected to compact star clusters and small associations as their bound parts, but sub-structure characterizes also extended associations. As a consequence in the literature there is a strong overlap between associations and aggregates, as far as their morphology is concerned. This overlap reaches a degree, where the distinction between aggregates and associations (for example based only on their sizes) may not even make sense. This difficulty in classifying groupings as different objects based on their stellar and spatial characteristics is further discussed later in this section (see Sect.\,\ref{s:uniscale}).

\begin{figure}[t!]
\centerline{\includegraphics[angle=0,clip=true,width=1.\textwidth]{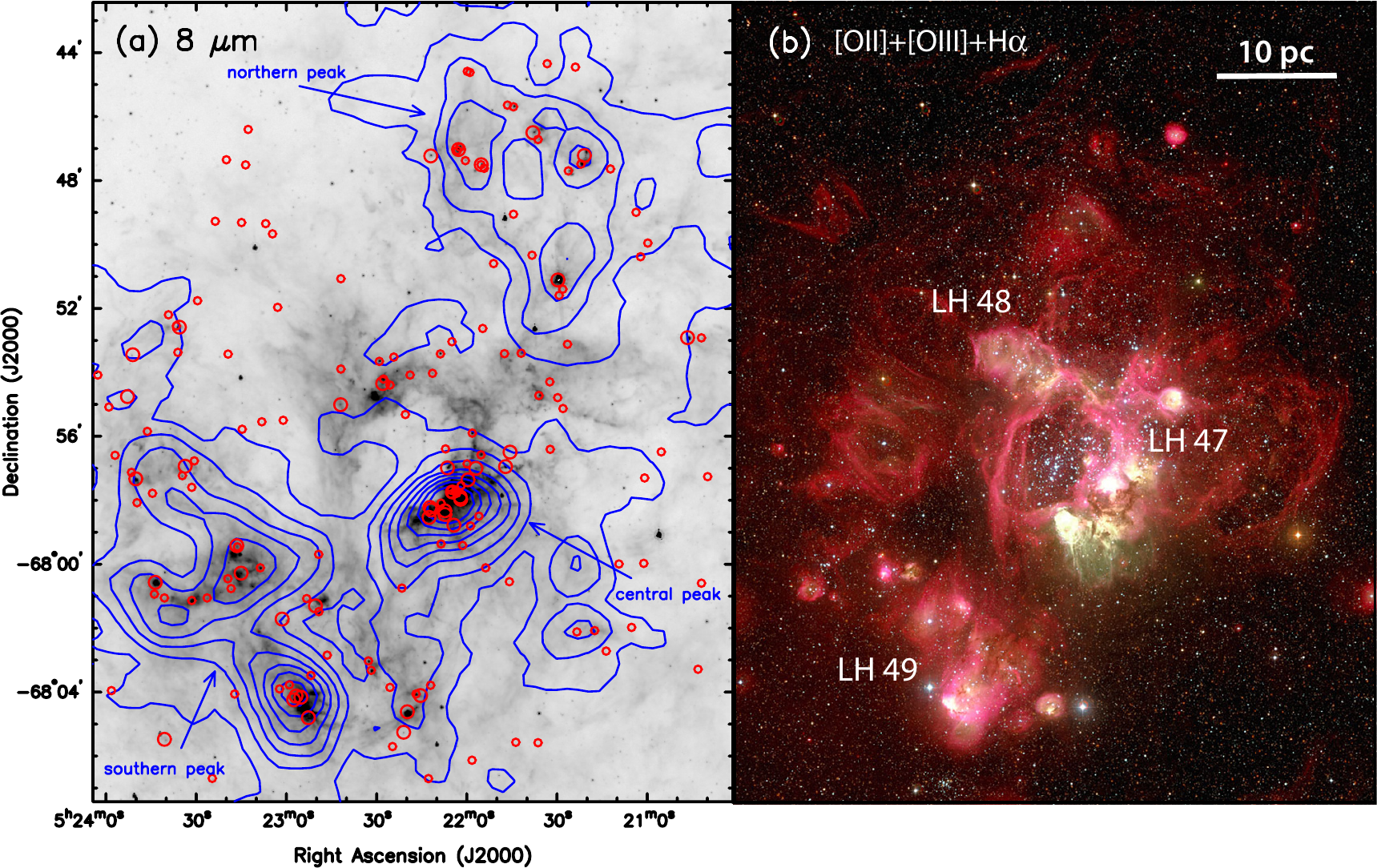}}
\caption{\small {\sl The star-forming complex N\,44 \citep{Henize1956} in the Large Magellanic Cloud}. (a) {\sl Spitzer} 8\,\micron\ image showing  PAH and dust emission, with CO contours \citep{Wong2011} overlaid in blue, and YSO candidates \citep{Chen2009, Carlson2012} denoted in red; (b) Color composite of images from the WFI at the MPG/ESO 2.2\,m telescope in three narrow-band filters, [O$\;${\small{II}}\relax] (372 nm, blue), [O$\;${\small{III}}\relax] (500 nm, green) and H$\alpha$ (656 nm, red), with the OB associations in the region, LH\,47/48/49 \citep{LuckeHodge1970}, labeled.
The green color indicates areas that are particularly hot. Acknowledgements/Credits: Spitzer/CO/YSO combination (left): R. Gruendl (University of Illinois). ESO color-composite image (right): F. Comer\'{o}n and N. Delmotte (ESO).
\label{f:n44}}
\end{figure}

Stellar associations and aggregates typically coincide with {\sl giant star-forming regions} and with larger star-forming complexes. {\sl Stellar complexes} are sizeable, asymmetric stellar congregations with sizes extending from $\sim$\,few\,100\,pc to $\sim$\,few\,kpc (those of the latter sizes are usually called super-complexes), comprising gas, clusters, stellar associations and early-type stars with ages between $10^6$ and $10^8$ years \citep{Efremov1989}. Initially such structures in the Large Magellanic Cloud (LMC) were known under the name of `stellar constellations' \citep{McKibbenNailShapley1953} or `super-associations' \citep{Baade1963}. The term `star complex' was suggested to designate groups of Cepheids in the Milky Way, which are much larger than typical star clusters, and older than typical OB associations with stellar periods indicating ages between 30 and100 Myr \citep{Efremov1978b}. 

The region of a typical stellar complex under formation is shown in Fig.\,\ref{f:n44}, where the LMC star-forming nebula LH{$\alpha$}\,120-N\,44 \citep{Henize1956} with its contiguous cohort of HII regions, bubbles, and young stellar clusters exemplifies star formation occurring as a sequence of local `hotspots' inside a giant molecular cloud. The positions of the OB associations in the region \citep{LuckeHodge1970} are also indicated in the image. The massive stars of the association LH\,47, located in the super-bubble of N\,44, are the primary suspects for the expansion of the bubble \citep{OeyMassey1995}. X-ray observations  reveal 10$^6$ K gas heated by 
fast stellar winds and supernova explosions \citep{Jaskot2011}. The effects of stellar energy feedback and in particular along the 
western rim of the bubble, where star formation may have been triggered by its expansion, are also evident through its H$\alpha$  
and {\sl Spitzer} images \citep{Chen2009, Carlson2012}. {\sl Herschel} dust mass maps reveal a unique hierarchical interstellar medium (ISM)  structure across the complex \citep{Hony2010}, and CO surveys show that star formation activity arises from one coherent molecular cloud complex \citep{Fukui2001, Wong2011}. N\,44 is thus a multi-generation star-forming ecosystem, as it samples different phases of star formation within a single molecular cloud.

The existence of stellar associations, aggregates and complexes as true over-densities of blue stars is more obvious in external nearby galaxies  \citep[e.g.,][]{Ivanov1987}. Outside the Milky Way we have an overview of the topology of star formation, which is more clear than by looking through the gaseous Galactic disk, where the true structure can be hidden along the line-of-site. The identification of extragalactic UYSS (without the aid of kinematic information) is, thus, easier than in the Milky Way, allowing the clear distinction of young stellar concentrations against their surroundings \citep[e.g.,][and references therein]{Gouliermis2000}. However, the determination of their physical boundaries is a quite dubious task, and so whether these systems represent a {\sl characteristic} scale of star formation is a controversial question \citep[][]{Gouliermis2011}. In a study, using different observational material of the same extragalactic targets, \cite{Hodge1986} shows that the characteristics of the stellar systems classified by various authors as stellar associations depend on the identification criteria, the image scale, and even the distance of the hosting galaxy. Since then, various studies based on various datasets and criteria could not establish an explicit definition of a stellar association that would allow a proper comparison in different galaxies \citep[see, e.g.,][]{Kontizas1994, Ivanov1996, Bresolin1998, Brown1999}. In the following section (Sect.\,\ref{s:search}) a compilation of studies dedicated on the identification of UYSS across various galaxies is presented. Their findings are also discussed in terms of the possible characteristic scales in space and time these systems represent.

\subsection{The Search for Unbound Stellar Systems\label{s:search}}

The search for distinct concentrations of bright blue stars on galactic scales over the last 
$\sim$\,60 years has utilized a variety of methods and observational material, starting with 
identifications by eye on photographic plates \citep[e.g.,][]{Morganetal1952}, to computer algorithms on 
photoelectric or photometric data \citep[e.g.,][]{Ivanov1991}, and sophisticated
cluster analysis techniques on digitized catalogs of resolved stellar populations 
\citep[e.g.,][]{Kangetal2009}. While the detected concentrations of OB-type stars in different galaxies show on average 
a striking resemblance in their sizes \citep[][]{Gouliermis2011}, observational limitations 
and diverse techniques result to the identification not of different groupings, but of different {\sl significance levels} of the same stellar systems \citep[see, e.g.,][for a demonstration of selection effects in the constructed surveys]{Hodge1986}. So, while low resolution images were revealing 
the large stellar complexes in a remote galaxy, more advanced observations were allowing for resolving the more significant (denser) associations or aggregates, ``embedded" in these complexes.

An interesting example of the advancements achieved between different surveys is that of the so-called ``OB associations'' 
identified by eye on photographic  plates in M\,31 by \cite{vandenBergh1964}. More than 20 years later \cite{Efremovetal1987} resolved (by eye again) these systems into smaller stellar groups, which they refer to as the ``real O-associations''. These 
authors also reclassified open clusters found by \cite{Hodge1979} as O-associations, and the majority of 
van den Bergh's OB-associations as ``star complexes'', i.e., the larger constellations that  contain the associations. 
This point provides the first indication that the clustering of early-type stars across galactic scales cannot be trivially classified into distinct classes of unbound structures. Indeed, an automated  search was published almost 10 years later by \cite{Battinelli1996}, who applied a cluster analysis to identify structures which correspond to the 500 pc-sized complexes described by \cite{Efremovetal1987} and to test for the clustering relation between the observed structures. Another example is that of M\,33, where \cite{Ivanov1987} in their survey of associations and complexes note that the objects identified as stellar complexes coincide with the OB associations of identified in this galaxy by \cite{HumphreysSandage1980}.

Inconsistencies in the determined characteristics of the detected systems found even in the same environments between different studies are, thus,  naturally originated by differences in the observational material and methods used \citep[see, e.g.,][for a review]{Gouliermis2011}, but they also signify an undoubted fact: {\sl Loose stellar systems are not easily distinguishable because they are parts of the same hierarchical structure}. 
\cite{vandenBergh1964} makes a clear reference to the observational constrains in identifying the `units' of recent star formation in M\,31, by noting that ``associations in M\,31 appear to be five times as large as those in the Galaxy'', based on the survey of Galactic OB associations by \cite{Schmidt1958}\footnote{Schmidt's associations survey is based on previous surveys by \cite{Markarjan1952} and \cite{Morgan1953}.}. He further notes that ``this apparent difference [...] probably results from selection effects'', since ``in the Galaxy {\sl only the dense cores of associations can be detected} against the rich background of early-type field stars''\footnote{A situation that will likely be greatly improved by GAIA distances.}. This point demonstrates the strong natural link between complexes and associations, as the latter being the denser parts of the former. Indeed, \cite{Battinelli1996} in their revisit of the stellar complexes detected by \cite{Efremovetal1987} in the Andromeda galaxy with their friend-of-friend algorithm conclude that there is a hierarchical arrangement of the stars in these constellations. Nevertheless, the detection efficiency of their algorithm was 
limited to the length-scale of $> 100$\,pc, and therefore the assessment of hierarchy in M\,31 was qualitative. The quantification of hierarchy in stellar clustering, and its dependence on stellar parameters, such as age, was originally investigated in the studies of \cite{ElmegreenEfremov1996} and \cite{EfremovElmegreen1998}. The results of these studies and the topic of hierarchy in loose stellar clustering is addressed later in the section.



\afterpage{

\begin{table*}[t!]
\begin{threeparttable}
\caption{Surveys of unbound stellar systems in galaxies of the Local Group.}
\label{tab:asssur}
\begin{tabular}{lrrrrrcl}
\toprule
\multicolumn{1}{c}{(1)}& 
\multicolumn{1}{c}{(2)} & 
\multicolumn{1}{c}{(3)} & 
\multicolumn{1}{c}{(4)} & 
\multicolumn{1}{c}{(5)} & 
\multicolumn{1}{c}{(6)} & 
\multicolumn{1}{c}{(7)} &
\multicolumn{1}{c}{(8)} \\
\multicolumn{1}{c}{Galaxy} & 
\multicolumn{1}{c}{Distance} &
\multicolumn{1}{c}{Number} &
\multicolumn{3}{c}{Size (pc)}&
\multicolumn{1}{c}{Detection} & 
\multicolumn{1}{c}{Surveys} \\
\multicolumn{1}{c}{name} &
\multicolumn{1}{c}{(Mpc)} &
\multicolumn{1}{c}{ of objects} &
\multicolumn{1}{c}{ min} &
\multicolumn{1}{c}{ mean} &
\multicolumn{1}{c}{ max} &  
\multicolumn{1}{c}{ Method$^\dagger$}         & 
\multicolumn{1}{c}{ references$^\ddag$}      \\
 \midrule
Milky Way & & 62 & &148 & 603 & Eye &\cite{Schmidt1958} \\
LMC  & 0.05 & 122 &17& 77&  300&Eye &\cite{LuckeHodge1970} \\
         &    & 153 & 21& 86& 190&Cnt & \cite{Gouliermis2003} \\
         &    & 56 & 163& 408& 1526&Cnt & \cite{Livanou2006}$^a$ \\
         &    & 2941 &  6& 30& 2085&Eye & \cite{Bica2008} \\
SMC & 0.06 &  70 &18 & 77& 180& Eye & \cite{Hodge1985b} \\
         &  &    31 &50 & 90&270& Aut & \cite{Battinelli1991} \\
         &   &  260&2&61&868& Eye & \cite{Bica2008} \\
NGC 6822 & 0.48 & 16 & 48&163& 305& Cnt & \cite{Hodge1977} \\
         &   &  6 &   & 72&    & Aut & \cite{Ivanov1996} \\
         &   &  24   & 179  &   374  & 699   & Cnt & \cite{Karampelas2009}$^a$  \\ 
         &   &  47   & 18  &   76  & 239   & Cnt & \cite{Gouliermisngc6822} \\ 
         &   & 201  & 18  &  128  & 2203   & Cnt & \cite{Gouliermisngc6822}$^b$ \\ 
IC 1613  & 0.74 & 20 & 68&164& 485& Cnt & \cite{Hodge1978}  \\
         &   &  6 &   & 83&    & Aut & \cite{Ivanov1996}  \\
M 31   & 0.77 & 188 & 100& 477&1200&Eye & \cite{vandenBergh1964}$^c$ \\
         & & 210 & 20& 80&    &Eye & \cite{Efremovetal1987} \\
         &&   15 &   & 83&    &Aut & \cite{Ivanov1996} \\
         &    & 894 &23& 40& 577&Cnt & \cite{Kangetal2009} \\
M 33    & 0.91& 143 &   &250&    & Eye & \cite{HumphreysSandage1980}$^c$ \\
         & &  460 & 30& 77& 270& Eye & \cite{Ivanov1987} \\
           & &  55 & 200& 570& 1300& Eye & \cite{Ivanov1987}$^a$ \\
         &  & 289 &  6& 66& 305& Eye & \cite{Ivanov1991Ap+SS} \\
         &  &  41 & 10& 40& 120& 	Aut & \cite{Wilson1991} \\
         &  &  8 &   & 87&    & Aut & \cite{Ivanov1996} \\
Pegasus  &1.21 &  3 &   & 65&    & Aut & \cite{Ivanov1996}  \\
Sextans A&1.37 &3 &   & 93&    &Aut & \cite{Ivanov1996} \\
UGC 8091    &2.18 &   3 &   &114&    & Aut & \cite{Ivanov1996}  \\
NGC 2403 & 3.18& 88 &160&348& 600& Eye & \cite{Hodge1985a} \\
UGC 5336   &3.77 &  3 &   & 72&    & Aut & \cite{Ivanov1996}  \\
NGC 6503 & 4.99&244 &34 &200 &2317 & Cnt& \cite{Gouliermis2015LEGUS}$^b$ \\
NGC  3621& 6.75&  944&  30 & 158&  2780  & Aut & \cite{Drazinos2013}$^b$  \\
NGC  5457& 6.87&  710&  30 & 185&  2162  & Aut & \cite{Drazinos2013}$^b$  \\
NGC ~~925 & 7.94& 775 &  30 & 182& 2954   & Aut & \cite{Drazinos2013}$^b$ \\
NGC 1566 & 9.15&949&21 &163 &1856 & Cnt& \cite{Gouliermis2017LEGUS}$^b$ \\
NGC  3351& 10.0& 560 &  31 & 234& 2203   & Aut & \cite{Drazinos2013}$^b$  \\
NGC  2541& 11.5& 689 &  30 & 209& 3483   & Aut & \cite{Drazinos2013}$^b$  \\
NGC 7331 & 13.9& 142 &   &440&    &Eye & \cite{Hodge1986}  \\
NGC  4548& 16.0&  832&   40& 282&    2540& Aut & \cite{Drazinos2013}$^b$  \\
NGC 4303 & 16.5&235 &   &290&    & Eye & \cite{Hodge1986}  \\
\bottomrule
\end{tabular}
    \begin{tablenotes}
      \small
      \item $^\dagger$ Explanations of the detection methods: {``Eye'':} Detection by eye on photographic plates, films, or CCD images. {``Cnt'':} Identification on isodensity or isophotal contour maps based on star or photon counts. {``Aut'':} Automated friends-of-friends grouping algorithms on stellar catalogs.
      
      \item $^\ddag$ {Surveys of structures expressly characterized as {\sl Stellar} (or {\sl OB-}) {\sl Associations}, unless if  indicated otherwise.}
      \item$^a$ Surveys of structures explicitly characterized as {\sl Stellar Complexes}.
       \item$^b$ Objects detected across the whole dynamic range of the hierarchy (from small associations to super-complexes).
       \item$^c$ Objects mischaracterized as associations, and later classified as complexes based on higher resolution data.\\
    \end{tablenotes}
  \end{threeparttable}
\end{table*}

\clearpage
}



Concerning the detected UYSS and their characteristics, Table\,\ref{tab:asssur} summarizes all surveys of such systems available in the literature. This is an updated and improved version of Table\,1 in \cite{Gouliermis2011}.  The galaxies names and their distances (according to NED\footnote{NASA/IPAC Extragalactic Database; URL:  \url{https://ned.ipac.caltech.edu}}) are given in Cols. 1 and 2 of the table, and the numbers of identified objects in each survey are given in Col. 3. The most commonly assessed characteristic of the detected structures in almost all surveys is their sizes. The mean size of the identified systems is provided for all surveys in Col. 5 of the table, either as given in the literature, or calculated here from the original data (after being digitized, depending on availability). For example, the sizes of \cite{Schmidt1958} associations in the Milky Way were determined from their given apparent angular extension in the sky and their distances, as assumed in the survey. Minimum and maximum values of the systems sizes are also provided in Cols. 4 and 6 of the table, when available. The detection methods used for constructing the surveys and the references to the related studies are shown in Cols. 7 and 8, respectively.

Table\,\ref{tab:asssur}  includes all compiled surveys of loose blue stellar systems, classified either as associations or complexes, {\sl independently} of the criteria used in the individual studies. However,  as mentioned above, each study was using different criteria for the identification of different objects, and so this compilation exemplifies the inconsistency in the literature concerning the classification of loose systems. Moreover, there are differences between the measured average sizes (and size ranges) of objects that are identified as stellar associations in different surveys, even for the same galaxy. The importance of these differences in the sizes of structures of the same class emanates from the assumption that {\sl stellar associations represent a universal length-scale of star formation} across various galactic environments.  This idea originates from \cite{Efremov1989}, 
who notes that OB-associations identified in the Galaxy, M\,31, M\,33, and the LMC have remarkably similar average diameters of 80 - 100\,pc, so that ``we can talk in terms of a {\sl universal} average diameter of O-associations in all galaxies''.  {The galaxy-to-galaxy similarity in the average diameters of OB associations led \cite{Ivanov1996} to suggest their use as a standard for computing galaxy distances.} \cite{Efremov1989} argues that also  stellar complexes, which comprise OB-type and later supergiants, may represent another universal length-scale  of star formation of  0.5 - 1\,kpc.  The issue of universal scales of star formation is an interesting topic, which is addressed in the following section (Sect.\,\ref{s:uniscale}) in view of the results of the surveys of Table\,\ref{tab:asssur}.

The average sizes of loose stellar systems in Table\,\ref{tab:asssur} agree well with the scale suggested for associations (with diameters between 70 and 90\,pc) only for half of the surveys of objects classified as associations. This count does not consider the 14 surveys, which focus on stellar complexes, those of objects found at various levels of the hierarchy (from small associations to large complexes), or those proven to include complexes misclassified as associations (see notes in the table). In the remaining half of the associations surveys, only four include objects with a mean size less than 70\,pc, while eight include associations with sizes larger than 90\,pc on average. One may argue that this inconsistency is due to the fact that in these surveys stellar complexes, which are typically larger than associations, have been misclassified as associations, as is the cases of M\,31 and M\,33, discussed above. While this may be the case, a detailed look at the members of the most recent surveys, which are complete in stars and more accurate because of their self-consistent methods, reveals a more complicated and interesting picture about the {\sl typical} length-scales of star formation.

\subsection{A Universal Scale of Star Formation? \label{s:uniscale}}

\begin{figure}
\floatbox[{\capbeside\thisfloatsetup{capbesideposition={right,top},capbesidewidth=0.5\textwidth}}]{figure}[\FBwidth]
{\caption{\small {\em Different types of stellar associations in the LMC}.
Size density distributions for the four sub-classes of associations identified in the LMC by \cite{Bica2008}. While the size of all objects classified as associations {\sl being related to nebular emission} averages around 85\,pc, the average size of all associations in the survey is far smaller and equal to about 30\,pc, due to the inclusion of resolved small associations. Typical sizes of objects in the individual sub-categories also differ significantly from one class to the other. The values of these sizes per sub-class are 13\,pc (AC), 33\,pc (A), 63\,pc (AN), and 160\,pc (DAN). For comparison, the corresponding sizes for associations in the SMC from the same survey are 18\,pc (AC), 59\,pc (A), 67\,pc (AN), and 90\,pc (DAN), which gives a total average larger than that for the LMC, of about 61\,pc. Given these distributions one may question the notion of a single characteristic size for stellar associations. {The size distribution of GMCs in the LMC \citep[][white transparent distribution with dashed line]{Wong2011} is also shown for comparison.}}\label{f:bicasizes}}
{\includegraphics[width=0.45\textwidth]{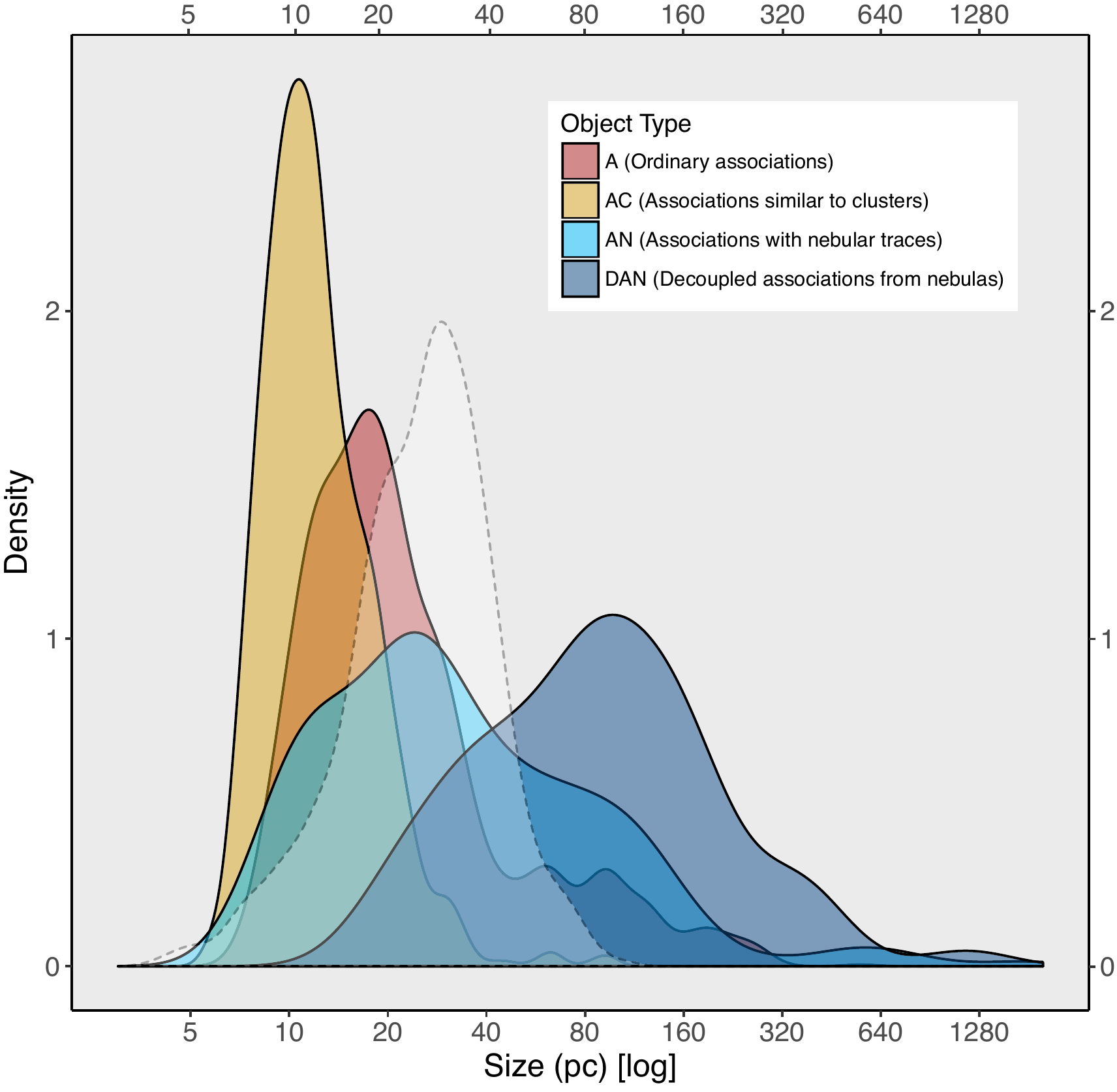}}
\end{figure}

For the analysis presented in this section let us consider the example of the surveys of stellar associations in the Magellanic Clouds, constructed by 
\citet[][see also \citealt{Bica1995} and \citealt{Bica1999} for the original compilations of the catalogs]{{Bica2008}}. These surveys refer in Table\,\ref{tab:asssur}  to all objects characterized by \cite{Bica2008} as associations, a class which in fact covers five individual sub-categories of systems defined by these authors: (1) ordinary associations (A), (2) associations similar to clusters (AC), (3) associations with nebular traces (AN), (4) decoupled associations from nebulas (DAN), and (5) nebulas with embedded associations (NA)\footnote{Note that there are no 
objects of class NA neither in the Small nor in the Large Magellanic Cloud catalog.}. The mean sizes and size ranges stated in Table\,\ref{tab:asssur} correspond to members of all these categories, and therefore they are the result of {\sl averaging over a range of morphologically different stellar systems}, which nevertheless belong to the same class of systems. Indeed, the detailed look at the size distributions of each of the associations sub-classes reveals objects with quite different dimensions (Fig.\,\ref{f:bicasizes}). If for instance we consider only nebula-related associations (in practice sub-classes AN and DAN), the average dimensions of these systems are 85\,pc in the LMC -- in  extremely good agreement with the suggested universal scale -- and 102\,pc in the Small Magellanic Cloud (SMC). If, however, we include ordinary small associations (A) and associations similar to clusters (AC), the average size of associations in the LMC  lowers to 30\,pc. This scale is not unrealistic, given that small, asymmetric, partially embedded associations are quite common in this galaxy \citep{Gouliermis2003}.


The survey of extended stellar systems in the Magellanic Clouds is the product of eye identification based on Sky Survey plates and the compilation from numerous previous catalogs sparsely distributed in the literature \citep[][]{Bica1995, Bica1999}. The categorization into several sub-classes demonstrates the large dynamic range in characteristics of systems of the same type. This and the large differences in the average sizes of typical associations between the Magellanic Clouds indicate that {\sl there are differences in the typical lengths of stellar associations between members of different sub-classes in the same galaxy, as well as between objects of the same sub-class in different galaxies}. This fact contradicts the universality of stellar associations dimensions as indication of a unique scale of star formation. 
Similar analyses of the data provided from other surveys of UYSS shown in Table\,\ref{tab:asssur} lead more or less to the same conclusion.  In these surveys, stellar associations actually represent loose young systems of various morphologies. Under these circumstances we can argue that stellar associations are not a single well-constrained type of unbound stellar system, but represent a part of a whole spectrum of young systems at various length-scales and with various degrees of gravitational self-binding \citep[see also][]{Elmegreen2000PPIV}.  {The size distribution of GMCs in the LMC occupies the middle ground in Fig.\,\ref{f:bicasizes} (white transparent distribution), implying that the smallest size associations probably originate from sub-structure within GMCs, while the largest objects are likely dynamically expanded associations. The GMC size distribution compares with, but it is not entirely similar to, that of the ordinary or nebula-related associations, and it peaks at almost the same scale with the latter.}

The length-scale across which star formation {\sl typically} occurs in galaxies is set most probably by the dimensions of the progenitors of UYSS, i.e., the star-forming GMCs (see, e.g., Fig.\,\ref{f:n44}). While this scale may be {\sl characteristic}, it is not necessarily universal, i.e., it is not independent of local conditions. {Galaxies in the Local Group demonstrate a diversity of star formation locations and conditions, which may explain the observed variations in the star formation length-scale.} Specifically, this scale should be naturally constrained by that of gravitational instabilities in star-forming structures. For example in spiral galaxies there are two such physically-driven scales \citep[see, e.g.,][for a review]{Elmegreen2011SFSpiralArms}. The first, studied by \cite{Jeans1902}\footnote{Jeans' analysis was further generalized by \cite{Chandrasekhar1955} and \cite{BelSchatzman1958}.}, addresses the balance between self-gravity and thermal pressure. The second scale results from the analysis by \cite{Toomre1964}, originally studied by \cite{Safronov1960}, which considers also shear forces from differential rotation in addition to gravity and pressure. In a galaxy with a global stellar spiral density wave or a stellar ring, large scale dynamics lead to gaseous structures like the so-called {\sl beads on a string} and {\sl spurs} \citep[e.g.,][]{Renaud2014, Schinnerer2013PAWSI}. The former are the result of gaseous gravitational instabilities, while the latter originate from Kelvin-Helmholtz instabilities. These structures should define the characteristic scale where stars form in a galaxy. If the galaxy has no spiral density wave, then instabilities in both stars and gas form the so-called ``flocculent arms" \citep{ElmegreenD1981}. These, so-called {\sl swing-amplifier}, gravitational instabilities  \citep{Toomre1981}, generally enhanced by magnetic fields \citep{KimWT2002}, may also drive at large scales turbulence \citep[see also][for a review]{Dobbs2014PASA}. Turbulence is a scale-free process known for producing self-similar gaseous structures and clouds \citep[see][for complete reviews on interstellar turbulence]{ElmegreenScalo2004, ScaloElmegreen2004, MacLowKlessen2004}. The formation of UYSS is observed to follow a similar scale-free clustering pattern \citep{Elmegreen2011Hierarchies}. This is discussed in more detail in the following section.

\greybox{UYSS cover a wide dynamic range in length-scales and compactness. Stellar associations and stellar aggregates represent a part of the whole spectrum of loose young systems. Their sizes are related to those of star-forming molecular clouds,  apparently shaped by galactic dynamics. The degree of gravitational self-binding of UYSS appears to also vary. A wide hierarchy in length- and time-scales continues from small embedded associations to gigantic super-complexes across galactic disks. We can thus assume that there is no universal scale of star formation, but the characteristics of the `unit' structure of star formation in galaxies are rather set by local (GMC-scale) and global (disk-scale) environmental conditions.}

 

\section{Hierarchical Young Stellar Clustering \label{s:hierarchy}}

\begin{figure*}
\includegraphics[width=\textwidth]{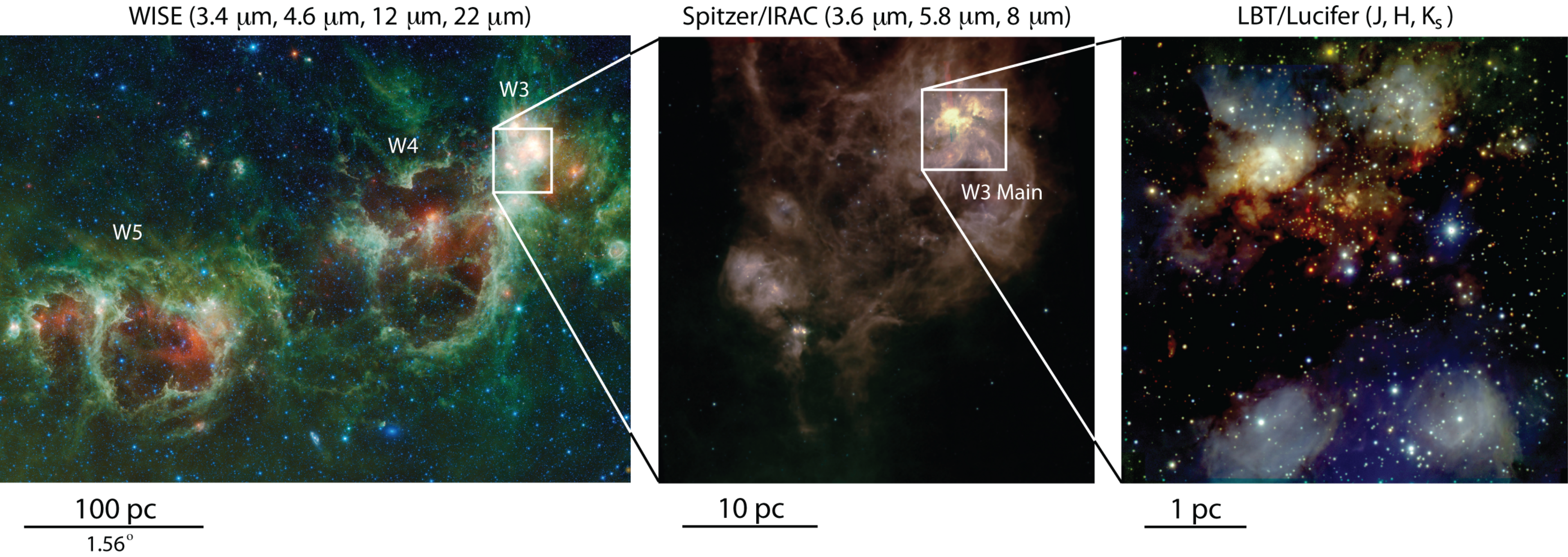}
\caption{\small {\em The hierarchical structure of the Interstellar Medium and Star Formation}.
Hierarchy in the general area of the Galactic W3/W4/W5 cloud complex. {\sl Left} -- The gaseous structures W4 and W5 (the heart and soul nebulae) are shown in the NASA/WISE infrared image, covering at 100-pc scale the general region of the OB association Cas OB6. {\sl Middle} -- The giant star-forming region W3, which is part of the W4 complex, is shown at 10-pc scale in the NASA/Spitzer image. The star-burst of W3 Main, highlighted by its bright PAH emission at 8 $\mu$m, is the most active part of W3. {\sl Right} -- The star-forming cluster of W3 Main is resolved in near-IR images from LBT/Luci at 1-pc scale. Acknowledgements/Credits: WISE image: NASA/JPL-Caltech/WISE Team. Spitzer image: The author. LBT image: \cite{Bik2012}. Adapted from \cite{Gouliermis2015PHAT}.
}\label{f:hierarchyw345}
\end{figure*}

Small stellar systems are typically parts of larger and sparser young systems, {which are not produced only by cluster dissipation, but they may also form in an unbound fashion at the peripheral regions of clusters (see, e.g., Sect.\,\ref{s:usf})}. Stellar structures extend, thus, across a wide range of lengths, from few pc cluster-like to kpc galaxy-wide scales. This hierarchical pattern in young stellar clustering, causing  -- as discussed above -- 
our difficulty in explicitly defining specific young groupings, comes naturally, given the fact that young loose systems are not distinct stellar systems isolated from their environments. In fact, it is known from early studies in the Milky Way and other nearby galaxies that stellar and gaseous structures are constructed in an orderly fashion (see, e.g., Fig.\,\ref{f:hierarchyw345}). Large stellar complexes are observed to include smaller more compact (but still loose) stellar associations, which are their brightest parts \citep[e.g.,][]{Ivanov1987}. Typical Galactic OB associations are known, since the beginning of last century, to be sub-structured themselves, with young star clusters being usually their compact parts \citep[see, e.g.,][]{Blaauw1964}. 

Sub-clustering within giant star-forming structures is confirmed in resolved stellar and cluster complexes in the Galaxy and Magellanic Clouds. These complexes are found to comprise smaller stellar groupings, which themselves ``break'' into smaller associations and clusters \citep[e.g.,][]{Gouliermis2014, Kuhn2014, DaRio2014}.  This morphology appears to be {\sl self-similar}, comparable to that of the interstellar molecular gas  \citep[e.g.,][]{Dickman1990, ElmegreenFalgarone1996}. Star formation, thus, seems to follow the same "architecture", which is often referred to as {\sl fractal}. Originally, fractal structure as a general property of interstellar gas was suggested by \cite{Scalo1985,Scalo1990}, and verified in various studies \citep[see, e.g.,][and references therein]{Falgarone1991}. The {\sl scale-free} structure of the interstellar medium is believed to be shaped mainly by turbulence \citep{Larson1981, Stutzki1998}, as predicted in laboratory and theoretical experiments on turbulent fluid dynamics \citep[e.g.,][]{Sreenivasan1991, Federrath2009}.  Fractals are shapes characterized by symmetries, which are invariances under dilations or contractions \citep{Mandelbrot1983}. The term "self-similar" is used in the sense that each part of the fractal shape is a linear geometric reduction of the whole, with the same reduction ratios in all directions\footnote{This term is extended to include self-affine shapes, in which the reductions are still linear but the reduction ratios in different directions are different.} \citep{Mandelbrot1989}. This section focuses on the connection between the scale-free nature of star formation and the hierarchical clustering of young stars. Evidence of this clustering pattern in various stellar environments of the Local Group is also presented.

\subsection{Clustering on Galactic Scales\label{s:clus@Glx}}

Young star clusters in the Milky Way and other local galaxies are organized in distinct groups when age, spatial distribution, and kinematics are considered. These structures are known as {\sl cluster complexes}, being identified by stellar clusters as their distinct members. However, they are also typically populated by plethora of young sparsely distributed stars, i.e., they are {\sl stellar complexes}. The differentiation between clusters and stars as members of these structures has introduced two separate paths in quantifying the clustering behavior of large-scale stellar structures: The investigation of groups of distinct star clusters, where only compact clusters are considered to build up a complex, and that of the extended distributions of young stars along a complex, where star clusters may or may not also considered (without a significant statistical impact on the results). 

\subsubsection{Cluster Complexes}

The majority of Galactic open clusters (OCs) are known to be members of groups with characteristic sizes of several hundred parsecs \citep{Eigenson+Yatsyk1988}. These groups are identified on the basis on the similarity of the ages, the integrated colors, and the radial velocities of the OCs. These criteria allowed to distinguish physically related OC groups  from random distributions, assuming common origin of the OCs that belong to the same group. Young OCs appear to be more clustered,  although the traces of a common origin in many of these groups are preserved for several hundred million years. Neighboring groups  could be identified as members of structures of higher order with sizes of up to few kpc. \cite{Eigenson+Yatsyk1988} distinguish 54 such groups that contain 199 OCs. An updated, more complete, catalog of Galactic OCs in the Solar neighborhood was compiled from Tycho-2, Hipparcos and other catalogs by \cite{Piskunov2006}. Fluctuations in the spatial and velocity distributions of this OCs population are attributed by these authors to the existence of four so-called "open cluster complexes" of different ages, identified on the basis of the same kinematic behavior and a narrow age-spread among their members. The youngest of these complexes is considered to be a signature of Gould's Belt, and the oldest complex ($\log{\rm age} \simeq 8.85$) is also the sparsest in the sample. In another study,  \cite{delaFuenteMarcos+delaFuenteMarcos2008} identified five dynamical families of OCs in the Solar neighborhood by using age- and volume-limited samples from the Open Cluster Data-base\footnote{WEBDA Database; URL: \url{http:// www.univie.ac.at /webda/}}\citep{Mermilliod+Paunzen2003}. These families -- named in order of distance, Orion, Scutum-Sagittarius, Cygnus, Scorpius, and Cassiopeia-Perseus -- are found to be associated to the underlying Galactic spiral structure. They are probably bound (at least initially) and the likely progenitors of moving groups and stellar streams. 

Clusters in these groups are also found to be grouped in hierarchical fashion \citep{delaFuenteMarcos2009}. The hierarchical distribution of young clusters suggests that dense, crowded environments at small scales are more efficient in forming clusters. This increases the possibility that clusters in binary or multiple systems are formed together, as theoretically predicted \citep{FujimotoKumai1997, Priyatikanto2016}. Probable binary cluster systems have been observed in the Milky Way \citep{Subramaniam1995}, and the Magellanic Clouds \citep{BhatiaHatzidimitriou1988, HatzidimitriouBhatia1990, Dieball2002}. Concerning the survival of Galactic cluster complexes, the groups of open clusters in the Milky Way are considered short-lived. {A sharp decline in the age distribution of young OCs within 2.5\,kpc from the Sun suggests a timescale of 20\,Myr for most (50\%--80\%) newly formed OCs to dissolve in the field \citep{delaFuenteMarcos+delaFuenteMarcos2008}.} This result and the high destruction rate of young clusters suggest that any cluster-related structure should be younger than about 30\,Myr, although coherent structures of OCs as old as at least 100\,Myr are observed to be dynamically induced \citep{delaFuenteMarcos+delaFuenteMarcos2008}. 

The clustering of star clusters and its dependence on age is investigated also in other local galaxies. While the autocorrelation function of star clusters of all ages in the M\,51 disk shows no obvious structures, the spatial distribution of clusters in three separate age ranges show that the youngest sample forms more crowded substructure, following a hierarchical pattern 
\citep{Scheepmaker2009}. These clusters are found inside groups, which are inside cluster complexes and so on, up to scales of $\sim 1$\,kpc. Young clusters in the Antennae galaxy are also found to be auto-correlated up to these length-scales \citep{Zhang2001}. The systematic increase of the clustering strength with decreasing star cluster ages was confirmed through star cluster correlation functions also in seven star-forming galaxies in the Local Group \citep{Grasha2015LEGUS, Grasha2017LEGUS06}. The distribution of star clusters was found to become more homogeneous after a timescale of $\sim$\,40--60\,Myr, and at a lengthscale larger than a few hundred parsecs. This timescale is in good agreement to that estimated for cluster complexes in the Milky Way \citep{delaFuenteMarcos+delaFuenteMarcos2008, delaFuenteMarcos2009}. {The corresponding mean velocity dispersion for the clusters to become distributed uniformly is also consistent with the velocity 
dispersion of field OB stars in the Solar neighborhood \citep[e.g.,][]{Gontcharov2012}.}  In all studied galaxies compact associations were found to be more strongly correlated than star clusters, which appear to move significantly away from their birth place within a few 10\,Myr. These studies suggest that the survival of young star clusters depends on the environmental conditions present in the star complex where they formed, as has already been pointed out by various investigations \cite[e.g.,][]{Hodge1987, Boutloukos2003}. 

\subsubsection{Stellar Complexes}

Stellar complexes are large constellations of young stellar populations extending from few 100\,pc to kpc scales \cite[][]{Efremov1989, Efremov2015A&AT}. As discussed in Sect.\,\ref{s:define}, they are the linking structures between associations and their parental galaxies. Recent studies focus on the identification of stellar complexes across a wide dynamic range of stellar densities and sizes. For instance, methods applied for the detection of stellar complexes use the length-scale or surface stellar density as free parameters. As a consequence the  constructed surveys include small systems within larger ones, revealing the hierarchical pattern in their structure (see Table\,\ref{tab:asssur} for details). \cite{Drazinos2013} in their study of six local galaxies apply a friends-of-friends algorithm  \citep[see][for the original implementation of the method]{HuchraandGeller1982} on stellar catalogs within given search radii, set by the expected dimensions for associations, aggregates, complexes, and supercomplexes. \cite{Gouliermisngc6822, Gouliermis2015LEGUS, Gouliermis2017LEGUS}  apply a contour-based cluster analysis for the detection of asymmetric stellar concentrations on surface stellar densities produced with star counts. The detection takes place in steps defined by the standard deviation of the overall density across the observed field \citep[see, e.g.,][for an original implementation]{Gouliermis2000}. 

\begin{figure*}[t]
\centering
\includegraphics[width=\textwidth]{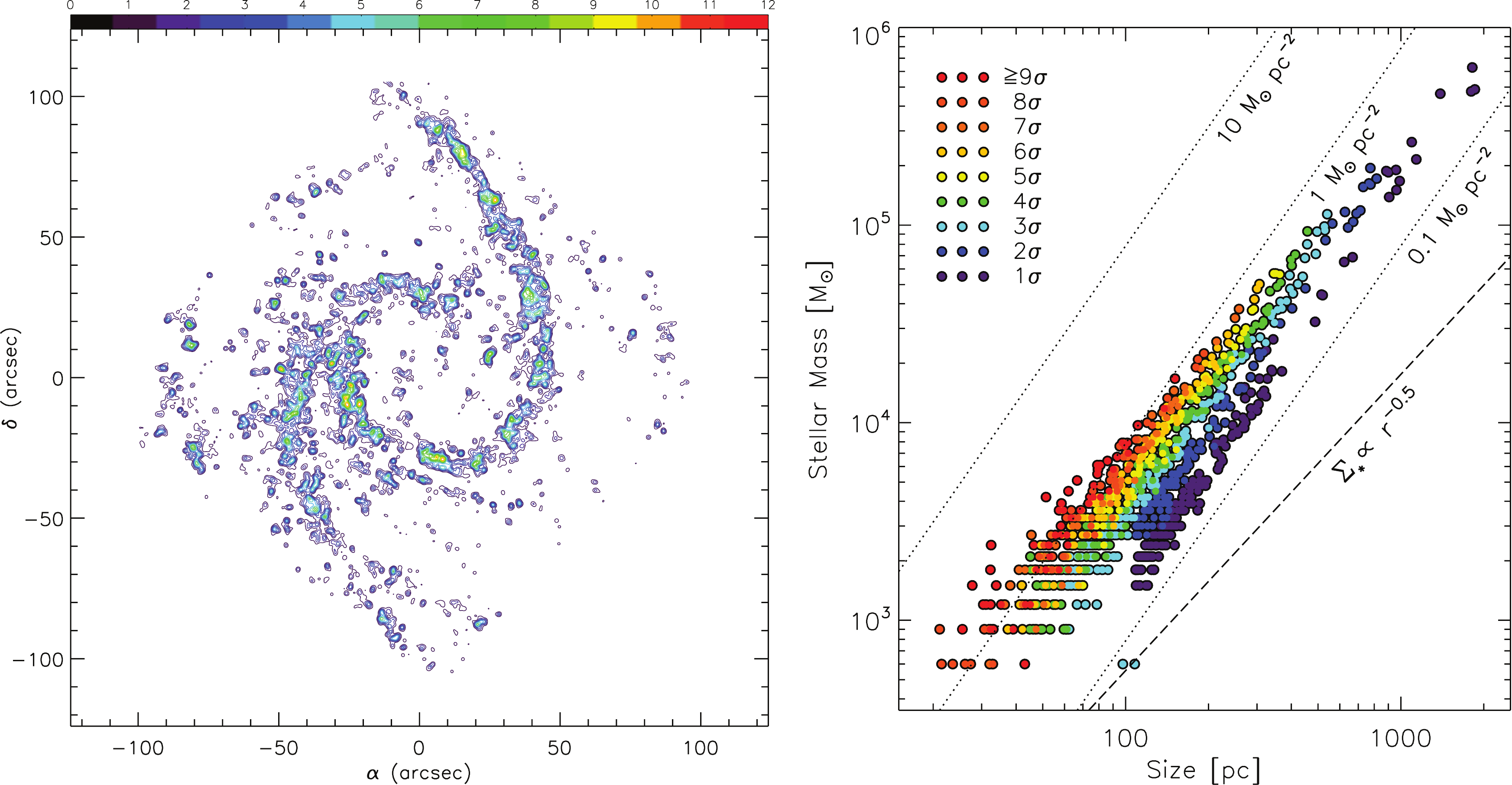} 
\caption{\small {\em Complexes within complexes across the grand-design spiral NGC\,1566}. 
{\em Left panel.} Surface number density map of the massive young stars in the galaxy. 
Different colors correspond to different significance levels used for the detection of stellar complexes across the galaxy. 
Significance levels are defined in standard deviations of the background density, $\sigma$, as indicated by the color bar.
The vast majority of star formation takes place in a hierarchical fashion along the spiral arms. Smaller (more compact) stellar structures are identified within progressively larger and more loose structures. {\em Right panel.} Mass--size relations for the stellar complexes detected at difference levels, as indicated in the legend. Structures found at low levels are the larger and looser in the sample. 
They have mass--size relations of almost uniform stellar distributions, as indicated by the dotted lines that show these relations for uniform stellar densities of 0.1, 1, and 10 M{\solar}\,pc$^{-2}$ respectively. Stellar systems identified at higher significance levels (i.e., smaller and more compact systems) show mass-size relations closer to these of condensed stellar systems, as indicated by the dashed line, which demonstrates the mass--size relation of centrally concentrated systems with a power-law radial stellar density profile of the form $\propto r^{-0.5}$. Power-law fits 
to these relations exemplify the deviation from uniform stellar super-complexes to condensed stellar aggregates, and quantify the 
hierarchy in the formation of these structures. {It should be noted that due to the minimum size-limit for the detection of stellar complexes the sample completeness introduces a bias against systems in the low-mass low-size part of the mass--size relation.} Adapted from \cite{Gouliermis2017LEGUS}.  \label{f:ngc1566}}
\end{figure*}

The constructed surveys of loose stellar systems include a wide range of structures, connected to each other through a parent-child relation in the sense that smaller (and more compact) groupings (e.g., associations) are members of larger (and looser) systems (e.g., aggregates), which themselves are the `compact' parts of even larger structures (e.g., complexes), which in turn `merge' into huge patches of young stellar conglomerations (usually termed ``super-complexes''). These studies show that, as discussed in Sect.\,\ref{s:uniscale}, the distinction of different levels in the hierarchy into separate classes of systems probably has no physical meaning, because the observed spectrum of stellar complexes in compactness and length-scale is not distinct, but continuous (see Fig.\,\ref{f:ngc1566}). The characterization of this hierarchical structure pattern as self-similar or fractal is based on the size distributions of nested star-forming complexes and the two-point correlation function of young stellar populations, both demonstrating power-law shapes \citep[e.g.,][]{Gouliermis2015LEGUS}. The exponents of these laws provide direct measurements of the {\sl fractal dimensions}\footnote{In fractal geometry, fractal dimension is a ratio of a statistical index of complexity comparing how detail in the fractal pattern changes with the scale at which it is measured \citep[e.g.,][]{Falconer2003}.} of the observed patterns. In fractal geometry a self-similar shape may be split up into $N$ parts, obtainable from the whole by a {\sl similarity} of ratio (i.e. scaling factor) $\epsilon$. The fractal dimension of such a shape is  defined as $D = \displaystyle -{\log{N}}/{\log{\epsilon}}$ \citep{Mandelbrot1983}. The interpretation and comparison of the measured fractal dimensions for both stellar and gaseous ensembles requires its own dedicated analysis, and it is therefore, outside the scope of this paper. It is worth mentioning, though, that the derived values are found to be very sensitive to the method used for their calculation \citep{Federrath2009}. Therefore, comparison of results derived from different methods are not unambiguous. Also, in contrast to the geometrical dimensions, the difference between the three-dimensional, $D_3$, and two-dimensional, $D_2$, fractal dimensions {\sl is not always one} (i.e., $D_3 \neq D_2 +1$), as  falsely reported occasionally in the literature  \citep[see, e.g., discussion on the issue in][]{Gouliermis2014}. 

\subsubsection{Survival of Structure with Time}

Theoretical predictions for star clusters to become partly unbound after gas expulsion by various feedback processes suggest a timescale of the order of 10\,Myr \citep[see, e.g.,][and references therein]{Pfalzner2014}. Rapid cluster evaporation feeds the hosting complexes with more dispersed stellar members, eventually influencing their stellar demographics and internal dynamics. The survival, however, of these complexes as spatially coherent structures apparently lasts for an order of a magnitude longer timescales. This is suggested by the structures of Cepheid variables (with ages up to $\sim$\,100\,Myr) found in both the Milky Way and the Magellanic Clouds \citep{Efremov1989}, and the estimated lifetime of Galactic open clusters of $\sim$\,300\,Myr \citep{Piskunov2006} and their structures of over 100\,Myr \citep[e.g.,][]{delaFuenteMarcos2009}. The clustering length-scale of young stars is found to depend on their ages, indicating that both star formation {\sl and} dynamics shape the geometry of star-forming structures with time. So, {\sl the evolution of stellar clustering with time at various scales is due to both nature (star formation) and nurture (dynamics)}. 

Concerning nature, the formation of stellar structure seems to proceed hierarchically {\sl in both space and time}. 
The duration of star formation tends to increase with the size of the region as the crossing time for turbulent motions.  This suggests that small regions form stars quickly and large regions, which contain the small ones, form stars over a longer period. This correlation underscores the perception that both, cloud and stellar structure, come from interstellar gas turbulence and suggests that star formation in a molecular cloud is mostly finished within only $\sim$~2 to 3~turbulent crossing times \citep{EfremovElmegreen1998}. Concerning nurture, older stars (as well as clusters), apparently being dynamically removed from their natal systems, are found to be more widely distributed than the younger ones, which seem to be closer to their birthplaces. The derived scaling relations for stellar separations as a function of evolutionary age essentially measure the survival timescale of young stellar structures. 

In a study of the structural evolution of the SMC, \cite{Maragoudaki2001} found that while the old stellar population of the galaxy shows a regular and smooth distribution, the younger stellar component has a highly asymmetric and irregular distribution. This was  interpreted as evidence of the strong impact to the SMC morphology of its close encounter with the LMC 0.2 to 0.4 Gyr ago. {The study of star clusters and associations in the different ambient of the spiral M\,83 showed that local environment plays a key role in cluster disruption \citep{Silva-Villa2014}.} Larger scale distributions of stars with decreasing luminosity (increasing age) is being documented also for the spiral NGC\,1313, where 75\% to 90\% of the UV flux is found to be produced by stars outside the clusters  \citep{Pellerin2007}. In the dwarf irregular IC\,2574 20 out of 75 identified dispersed stellar groups were found to be  dissolving \citep{Pellerin2012}. If it is assumed that the dispersed stellar population is entirely originated in dissolving compact stellar groups, then there is a low number of such systems in IC\,2574, compared to the theoretical expectations \citep{BaumgardtKroupa2007}. This suggests that star formation in IC\,2574 possibly occurred mostly in unbound systems. This result is in line with observations in other galaxies, where the origin of the distributed young populations could not be entirely tracked to dissolved compact systems but also to star formation. 

\begin{figure}
\floatbox[{\capbeside\thisfloatsetup{capbesideposition={right,top},capbesidewidth=0.5\textwidth}}]{figure}[\FBwidth]
{\caption{\small {\em Survival of Stellar Structures across Time and Space in NGC\,6503}. The spatial evolution of large-scale stellar clustering as function of stellar age is depicted by the scaling relation between the average MST edge-length of young stars in eight equally-populated age intervals and the average age per interval. This plot is constructed with data from \cite{Gouliermis2015LEGUS}. The positive trend between length- and time-scale continues to MST edge-lengths up to about 80\,pc, corresponding to an average stellar age of $\sim$\,90\,Myr. The power-law correlation is extremely good with $R^2$ of about one. After these length- and time-limits apparently the relation flattens, but based only on one point due to the low-number statistics in the fainter (older) stellar age bin. Note the difference between the time-scale of $\sim$\,90\,Myr derived with this analysis and the somewhat shorter of $\sim$\,60\,Myr, derived  by \cite{Gouliermis2015LEGUS} based on the significant change in the slope of the autocorrelation function for the same stellar samples.}\label{f:lgthtime_evol}}
{\includegraphics[width=0.45\textwidth]{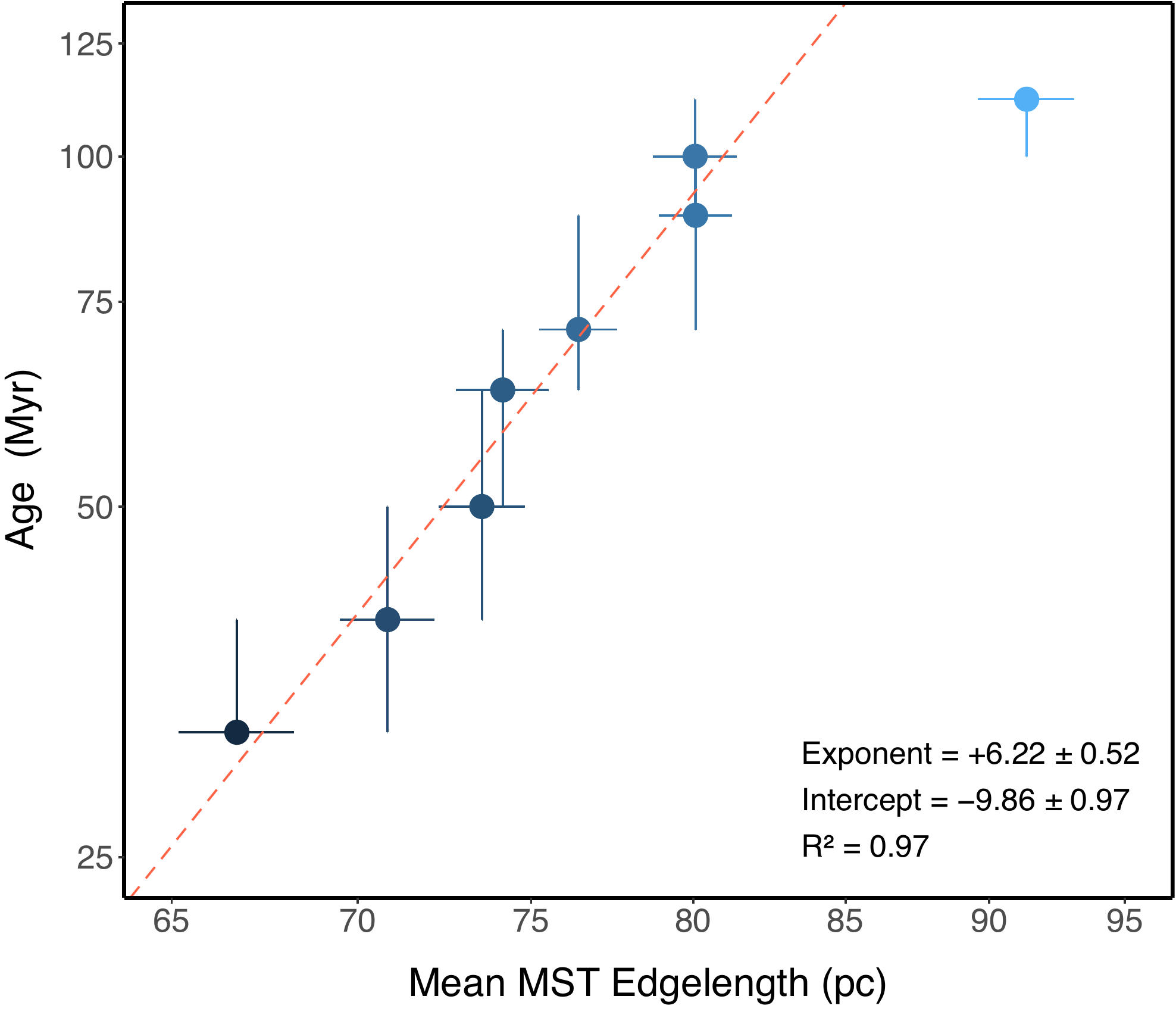}}
\end{figure}

The clustering behavior  of young stars at different evolutionary stages was investigated in the Magellanic Clouds \citep{Gieles2008, Bastian2009} and the ring star-forming galaxy NGC\,6503 \citep{Gouliermis2015LEGUS}. The autocorrelation function of stars in different equal-number luminosity intervals is found to be systematically steeper for the younger stars, corresponding to smaller fractal dimensions and more clumpy distributions than those for the older stars, suggesting that younger (brighter) stars are ``more clustered'' than the older. These findings were confirmed with the pair separations and minimum spanning tree (MST) edge-length probability distributions of the stars in various magnitude (age) ranges (Fig.\,\ref{f:lgthtime_evol}). The timescale, on which a significant change from a more clustered to a more distributed assembling of stars takes place, was determined to be $\sim$\,60\,Myr in NGC\,6503, $\sim$\,75\,Myr in the SMC, and $\sim$\,175\,Myr in the LMC. {The timescale for stellar diffusion is also determined to be of the order of $\sim$\,100\,Myr in the dwarf irregular NGC\,2366 \citep{ThuanIzotov2005} and the spiral NGC\,2403 \citep{Davidge2007}.} However, this time-scale, was found to be noticeably larger in M\,31, where young stellar structure apparently survives across the whole extent of the galaxy for at least 300\,Myr \citep{Gouliermis2015PHAT}, remarkably similar to the lifetime of open clusters in the Milky Way \citep[$322 \pm 31$\,Myr,][]{Piskunov2006}. This coincidence in timescales suggests that structure may dynamically survive for this long also in our Galaxy.




\subsection{Clustering on GMC Scales\label{s:clus@GMC}}

Multiplicity and sub-structure in young stellar concentrations comes as a natural effect of hierarchy. Star-forming regions in the Galaxy host 
significant sub-clustering down to sub-pc length-scales \citep[e.g.,][]{Schmejaetal2008}, where 
self-gravity \citep{Goodman2009Nature} and hierarchical gravitational fragmentation \citep{VazquezSemadeni2009fa} 
are considered to govern the physical conditions for low- and high-mass star formation.
At somewhat larger scales, compatible to those of typical GMCs, OB stellar associations comprise of individual small young clusters and 
sub-associations \citep[e.g.,][]{Brown1999ASIC, StahlerPalla2005}, while they themselves are members 
of stellar aggregates \citep[][]{Blaauw1964, Efremovetal1987}. These hierarchical systems show surface density distributions with multiple peaks, and significant structure over $\sim$\,100\,pc spatial scales. In contrast, centrally condensed clusters show highly-concentrated stellar distributions, and they have smooth radial profiles compatible with, e.g., simple power-law functions ($\rho_\star \propto r^{-\gamma}$), or King-Michie-like (isothermal) potentials \citep{King1962, Michie1962} over scales of few pc.

\begin{figure*}[t]
\centering
\includegraphics[width=\textwidth]{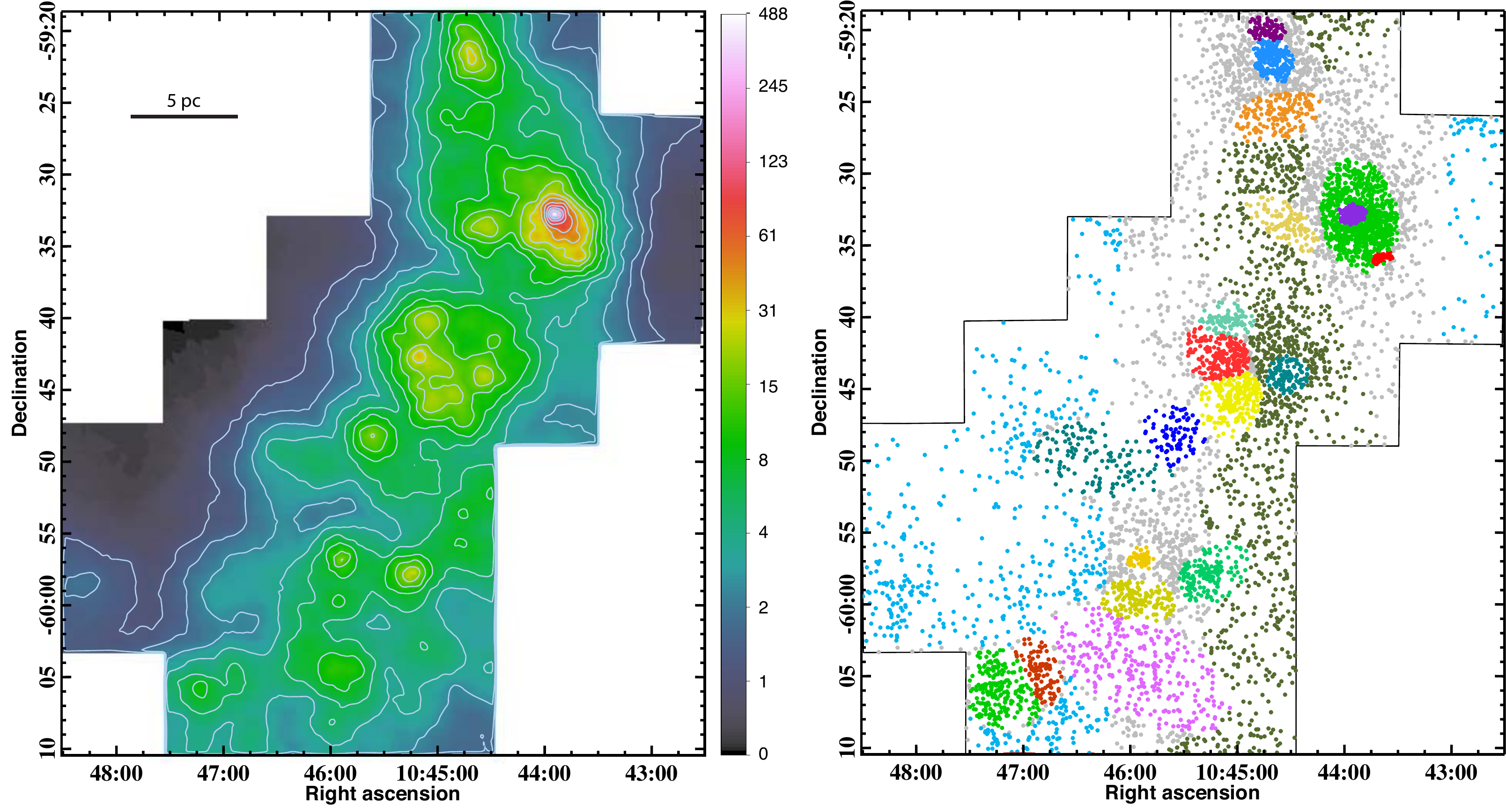} 
\caption{\small {\em Sub-clustering in the Carena Nebula star-forming complex with MYStIX}. 
{\em Left panel.} Smoothed projected stellar surface density map (color bar corresponds to units of observed stars per pc$^2$). 
Light-blue contours show increase in surface density by factors of 1.5. {\em Right panel.} The PMS stars in the Carina Nebula colored according to their sub-cluster assignments (20 sub-clusters). Light blue indicates the non-clustered stellar population, and gray the unassigned stars. Adapted from \cite{Kuhn2014}. Used with permission. \label{f:carinaneb}}
\end{figure*}

{\sl Hipparcos} data of nearby associations have shown that OB Galactic associations have stellar samples that extend to low-mass stars and consist of various subgroups and new associations \citep{deZeeuw1999}. {While from the structural point-of-view the evolutionary timescales of stellar associations and open clusters are very different, no clear-cut indication of such difference was found with these data.} Spatial substructure has been also identified in open clusters of the Milky Way as old as $\sim$\,100\,Myr \citep{SanchezAlfaro2009}. {The early formation of sub-structures in open clusters, and its subsequent evolution is investigated by \cite{Spera2016}, who interpret it as a consequence of the strong mass segregation, which leads to Spitzer's instability.} Observations of star-forming complexes and giant HII regions in the Galaxy appear to confirm that substructure is the result of hierarchical star formation across GMCs. The {\sl Massive Young Star-Forming Complex Study in Infrared and X-ray} \citep[MYStIX;][]{Feigelson2013} survey mapped the stellar content of 20 OB-dominated Milky Way star-forming regions at distances within 4\,kpc from the Sun. The spatial distribution of the MYStIX X-ray- and IR-excess-selected Pre--Main-Sequence (PMS) candidate stars (plus OB stars from the literature) showed that 17 of the surveyed regions have significant substructure. They confine ensembles of young sub-clusters, which are identified from the stellar samples as isothermal ellipsoids using finite mixture models \citep{Kuhn2014}. The richest region in clusters is the Carina Nebula star-forming complex, where 20 distinct sub-clusters are identified (Fig.\,\ref{f:carinaneb}). 

Based on the morphological arrangements of sub-clusters, the MYStIX survey reveals four classes of spatial stellar structure, i.e., long chains of sub-clusters, clumpy structures, single clusters with a core-halo structure, and isolated isothermal ellipsoid clusters \citep{Kuhn2014}. The  characteristics of the individual sub-clusters suggest significant dynamical evolution \citep{Kuhn2015}, with indications that young stellar clusters grow from hierarchical sub-cluster mergers \citep[see, e.g.,][]{McMillan2007}, and evidence of sub-cluster expansion, postulated as a consequence of gas expulsion due to stellar feedback \citep[e.g.,][]{Tutukov1978, Pfalzner2011}. Most of the sub-clusters appear not to be dynamically relaxed through two-body encounters. They should be also gravitationally bound for their structure to be preserved if their star-forming regions have velocity dispersions $\geq\,3$\,km\,s$^{-1}$ \citep{Kuhn2015}. Baring in mind that cluster analysis is sensitive to the method used for the identification of sub-clusters (MYStIX favors the detection of isothermal ellipsoids), the results discussed here draw a picture of multi-clustered dynamically active star-forming complexes, pretty similar to that derived from other investigations in the Galaxy and the Magellanic Clouds.

Using the MYStIX stellar survey, \cite{Jaehnig2015} investigated the degree of angular substructure in the stellar position distribution in relation to the distance from the sub-clusters centers, surface number density of stars, and local dynamical age. These authors found that the centers of the sub-clusters appear smoother than their outskirts, indicating a fast  dynamical processing on local dynamical timescales. Smoother distributions were also observed in regions of higher surface density, or older dynamical ages, suggesting that sub-structure was erased dynamically quite early or even while the cluster was still forming. An investigation of the Orion Nebula Cluster (ONC) with the same methodology provided similar results with the cluster's core appearing rounder and smoother than the outskirts, consistent with a higher degree of dynamical evolution \citep{DaRio2014}. This study showed that ONC is moderately supervirial due to the insufficiency of the observed mass to reproduce the observed velocity dispersion from virialized motions, indicating recent gas expulsion from the cluster.

The analysis of the spatial distribution of young stellar objects (YSOs) in the Orion Molecular Cloud Complex 
revealed several stellar clusters and groups as continuous regions with surface densities $\geq$\,10\,pc$^{-2}$ \citep{Megeath2016}.
All detected sub-clusters with more than 70 members exhibit asymmetric and elongated structure. In agreement with \cite{DaRio2014} findings, \cite{Megeath2016} confirm that ONC becomes azimuthally symmetric in its inner part (at 0.1\,pc), and has an unusually low star formation efficiency. 
Almost 60\% of the YSOs in the Orion Cloud Complex were found in ONC, while 13\% of the YSOs are distributed in a dispersed (non-clustered) fashion. The picture described above, of star formation taking place in both clustered and sparse distributions, is being also observed outside the Galaxy, specifically in giant star-forming regions of the Magellanic Clouds.

The SMC star-forming nebula N\,66 \citep{Henize1956} is found to host PMS stars, which are both grouped in various sub-clusters and dispersed across the complex, confirming previous findings that the star-forming complex exhibits considerable substructure \citep[e.g.,][]{Sabbi2007, Schmeja2009}. Based on simulated synthetic distributions the stellar clustering in N\,66 is explained as the combination of two separate components, a clustered and a non-clustered one, with $\sim$\,40\% of the observed PMS stars being located in centrally condensed systems, particularly in the young massive cluster NGC\,346, and the non-clustered stellar component to be structured in a self-similar pattern, 
spread across the whole complex \citep{Gouliermis2014}. This  is interpreted as the outcome of turbulent-driven hierarchical star formation, and substructure is considered to be due to the clumpy fractal distribution of the stars. 
No age-difference could be detected between the clustered and non-clustered populations. In a follow-up study that combined stellar photometry with dust emission maps, the SFE in N\,66, derived from star counts, is also found to vary significantly within the complex and to scale positively with surface stellar density \citep{Hony2015}. The highest SFE is measured at the position of the massive cluster NGC\,346, in agreement with theoretical expectations, according which SFE depends on the local gas density \citep{Pudritz02, Clark2007, KrumholzTan2007, PadoanNordlund2011,  Kruijssen2012, FederrathKlessen2013, Pfalzner2014}.

The multiplicity of local OB associations is quantified more accurately from kinematical studies. The Gaia-ESO survey resolved the velocity structure of the young (10\,-\,15\,Myr) cluster around the massive Wolf-Rayet binary system $\gamma^2$\,Velorum, in the Vela\,OB2 association, into two kinematic components \citep{Jeffries2014}. The first has an intrinsic dispersion consistent with virial equilibrium, while the second has a broader dispersion. The first stellar component is found to be 1 to 2\,Myr older, and is probably more centrally concentrated around $\gamma^2$\,Vel. A kinematically distinct component was also found in the young ($\sim$\,35\,Myr) cluster NGC\,2547, also in Vela\,OB2, which appears more similar to one of the  $\gamma^2$\,Vel components \citep{Sacco2015}. These studies depict a young, low-mass stellar population spread over several square degrees in the Vela OB2 association. This population may have been originated in a cluster around $\gamma^2$\,Vel that expanded after gas expulsion, or was formed in a less dense environment, spread over the whole association region.

Along the same lines, the Cygnus OB2 association is argued to have been ``always an association", in the sense that star formation took place in an non-clustered fashion with massive stars possibly born in relatively low density environments \citep{Wright2014}. From an X-ray selected sample of young stars, Cyg\,OB2 was found to host significant substructure with no evidence for mass segregation, both indications that the system is not dynamically evolved. The dynamical youthfulness of the association was confirmed with kinematical studies which suggest that the region is gravitationally unbound \citep{Kiminki2007, Wright2016} and therefore the massive stars in the association were never grouped closely together. Comparison with N-body simulations suggests that Cyg\,OB2 formed as a highly substructured, unbound association with a low volume density ($<$\,100\,stars\,pc$^{-3}$). More recently, {\sl Gaia} observations of populous unbound associations within 1.6 kpc from the Sun also do not show signs of significant expansion that would be expected if they would have formed as clusters that later expanded \citep{WardKruijssen2018}, suggesting that these systems were formed unbound.

\greybox{Galaxy-wide investigations of the young stellar component structure show that star-formation is organized in a self-similar manner, i.e., it is fractal over a wide range of scales. This fractal behavior in young stellar clustering may have been inherited by the ISM distribution, which is also self-similar. Hierarchical UYSS extend from the typical scales of star-forming GMCs up to those of spiral arms and they comprise both clustered and distributed populations of young stars and clusters. Their structure appears to scale in space and time due to both natural (star formation) and nurtural (dynamics) effects.}

There are competing theoretical models of how massive stars build up their considerable masses through processes such as those similar to low-mass star formation \citep[e.g.,][]{McKeeTan2003}, collisions or mergers in the cores of dense clusters \citep[e.g.,][]{ZinneckerYork2007}, or ``competitive accretion''  between originally low-mass molecular cores, some of them accreting considerable amount of matter due to their preferential positions in clusters' centers \citep[e.g.,][]{BonnellBV2004}. These scenaria, in particular the latter two, suggest as a condition of massive stars to form their location in the high-density centers of star clusters. The findings of \cite{Wright2014} and particular the result that it is highly unlikely that Cyg\,OB2 was ever a single compact cluster (currently in the process of dissolution due to post-gas expulsion) are inconsistent -- according to these authors -- with the notion of massive star formation theories that ``all stars form in dense compact clusters''. However, recent results on the search of isolated formation of massive YSOs in the LMC, suggest that massive stars (although not as massive as Cyg\,OB2 members) can form in isolated small compact embedded proto-clusters \citep[][]{Stephens2017}. Therefore, while the massive stars in Cyg\,OB2 may not have formed in the center of a massive cluster -- progenitor of the association, they may have formed within their ``own'' small compact systems spread across the star-forming molecular cloud in an unbound fashion. This is confirmed from the identified kinematic substructures of the association, which appear to be close to virial equilibrium \citep{Wright2016}. Sub-clusterings in unbound associations may start dissolving almost immediately after formation due to rapid gas dispersal \citep[e.g.,][]{MoeckelBate2010}, or because they are not gravitationally bound themselves. Such may be the case of 15 massive stars found in apparent isolation in the vicinity of the starburst 30 Doradus in the LMC \citep{Bressert2012}. While these stars are assumed to have formed in relative isolation, their radial velocities are similar (within 1$\sigma$) to the mean radial velocity of all observed massive stars in the region, suggesting that they are members of the same unbound structure.

\section{The Origin of Unbound Young Stellar Systems \label{s:usf}}

The studies discussed in the previous section draw a picture of ``clustered star formation in an unbound fashion'', in the sense that compact clusters form at the loci of dense gravitational centers across the star-forming complex. Star formation is thus more efficient in the gravitationally bound regions of the cloud, and while these regions may not be bound to each other, their dynamical evolution starts  already at very early stages. 
The outcome of this complex process is the formation of massive sub-structured UYSS, such as associations and aggregates.  Of particular interest are the initial conditions in the GMC that would favor the formation of such systems \citep[e.g.,][]{Krumholz2014PPVI, Padoan2014PPVI}. Young massive clusters (YMCs) may form in the densest regions of collapsing turbulent molecular clouds {\sl ``monolithically''}  after a free-fall time \citep[e.g.,][]{BallesterosParedes2015, FujiiPortegies-Zwart2016}, or through hierarchical clump merging \citep[e.g.,][]{Fellhauer2009, Smith2011, Kuznetsova2018} or by the assembling of less massive sub-clusters \citep[e.g.,][]{BanerjeeKroupa2015, Fujii2012}. The latter two scenarios imply a causal relation between the formation of clusters and unbound stellar systems. Indeed, as it is discussed later, bound cluster formation is also possible to take place within unbound molecular clouds. Therefore, while the production of UYSS may follow a different path than the monolithic formation of YMCs, both processes may share a common origin. Whether the formation of OB associations and other UYSS represents a different mode of star formation than that of bound and dynamically relaxed stellar clusters is still unclear. 


\subsection{Molecular Clouds as Progenitors of UYSS\label{s:GMCs}}

Star formation is governed by the localized gravitational collapse of cold molecular gas. GMCs in galaxies are the primary reservoirs of this star-forming gas, and thus understanding GMC formation, evolution, and destruction is essential in comprehending the physics of star formation. Fundamental quantities in our understanding of the physics of GMCs are their masses, areas, and velocity dispersions. Scaling relations between these quantities characterize the internal structure of the clouds, and signify the important role of gravity in setting the characteristics of clouds, either through collapse or virial equilibrium \citep[e.g.,][]{Ballesteros-Paredes2011a}. These scaling relations suggest a turbulent velocity spectrum, and a constant surface density for clouds \citep[e.g.,][]{Larson1981, Heyer2009}. GMCs constant surface densities have been confirmed for various galactic environments \citep[][]{Bolatto2008, Roman-Duval2010, Wong2011}, although they have been also challenged as observational artifacts \citep{BallesterosParedes2012}. In pressure-confined clouds the observed turbulent pressure $P = \rho \sigma_{v}^2$ (for density $\rho$ and velocity dispersion $\sigma_{v}$), is typically larger than the mean thermal pressure of the diffuse ISM \citep{Field2011, JenkinsTripp2011}, and so these structures are supported by self-gravity\footnote{Magnetic forces, while not a dominant factor, are considered important in balancing this self-gravity, depending on the  critical mass for a given magnetic flux threading it \citep{MouschoviasSpitzer1976}. 
}. On the other hand, hierarchical globally collapsing clouds, where kinetic motions originate from the collapse itself and not turbulence, may also appear virial  \citep{VazquezSemadeni2007}. Nevertheless, since in general self-gravity  is not an important condition for the gas in the clouds to be mainly molecular, GMCs should not necessarily be gravitationally bound \citep{Larson1993Ringberg}.

Molecular clouds form also without self-gravity due to galactic streaming motions, such as those induced by spiral shocks, and various sources of supersonic turbulence, like supernova explosions and other forms of feedback \citep[e.g.,][]{Dobbs2008, Bonnell2013}. Radiation pressure, cosmic rays, stellar winds, expanding HII regions, Kelvin-Helmholtz instabilities, and Rayleigh-Taylor instabilities are few other sources of supersonic turbulence \citep{ElmegreenScalo2004}. 
Earlier studies on star formation rates \citep{ZuckermanPalmer1974} and the transition from atomic species into molecules \citep{Jura1975} suggested that GMCs are long-lived structures. However, more recent investigations suggest that GMC formation and dispersal -- as well as star formation -- occurs on roughly one crossing time \citep{Ballesteros-Paredes1999A, Elmegreen2000ApJ}. Converging flows, producing compressions or shocks that force the gas over a critical density, lead to thermal instabilities and the formation of dense cold clouds. These supersonic shocks are highly dissipative, and without replenishment they disperse within a cloud crossing time.  Energy dissipation probably plays a determining role in the formation of filamentary structure, apparently a general feature of molecular clouds \citep[e.g.,][]{Andre2010,  Ward-Thompson2010, Arzoumanian2011}. Filaments are in general formed by the stretch induced by turbulence, and they survive longer in magnetized flows \citep{Hennebelle2013}. They usually made up of a network of short ribbon-like sub-filaments, and their existence seems to be unrelated to the boundness of the cloud \citep{Smith2014}. 

Scenarios of cloud formation consider GMCs to be highly molecular above a specific column density threshold or highly atomic below it \citep[e.g.,][]{Hartmann2001a, Krumholz2009a, Krumholz2009b}. According to these models, this threshold may arise from a shielding layer around the molecular part of a massive cloud, which implies a sharp decline in the radial profiles of the molecular fraction across galactic disks. Indeed these models explain the disk properties of spiral galaxies, but they do not seem to apply to dwarf irregular galaxies, which demonstrate smooth radial profiles \citep[][]{ElmegreenHunter2015}. This implies that there may be a difference between disk and dwarf galaxies in how GMCs are assembled. Galaxy-wide hydrodynamical simulations show that GMCs appear to be more often unbound than bound \citep{Dobbs2011}. {Concerning the lifetime of GMCs,  \cite{Murray2011} finds a characteristic lifetime of 27\,$\pm$\,12\,Myr, while \cite{DobbsPringle2013} find a lifetime of 4\,-\,25\,Myr.} Most of the clusters formed in such simulations are also unbound, because they cannot be followed over time for long before they disperse \citep{Dobbs2017}. Only clusters associated with long-lived, massive complexes are found to be bound. 

Self-gravity controls the formation of stars and clusters at small scales, but since these scales need not coincide with the scale of the thermal instability, much of the star-forming molecular cloud could be still gravitationally unbound \citep{Bonnell2011}. 
While self-gravity is shown to play a significant role in the internal structure of molecular clouds, it does not dominate the whole length-scale range in the clouds, becoming progressively less dominant at  larger scales \citep[][]{Goodman2009Nature}. Star-forming clouds with small, isolated regions of self-gravitating gas lead to {\sl distributed star formation}, i.e., star formation in an unbound fashion, at low efficiency. The formed stars should mostly have masses near their local Jeans mass. Regions of more gas, and hence more self-gravitating mass, form stellar clusters at a higher efficiency and populate a full stellar IMF, especially its high-mass regime \citep{SmilgysBonnell2017}. 
Indeed, the high-density cluster-forming regions show higher-than-average star formation efficiencies, leading to centrally concentrated stellar clusters \citep[e.g.,][]{Ginsburg2016}. An initially marginally bound stellar protocluster can also be easily disrupted if the conversion of cores into stars is inefficient \citep{KlessenBurkert2000}. 

Star formation efficiency (SFE) across a star-forming cloud seems to depend on gas density \cite[e.g.,][]{Pudritz02, PadoanNordlund2011, Pfalzner2014}. Stellar feedback limits the efficiency of the GMC in forming stars, but it is possibly a second order effect \citep{DaleBonnell2012}. The overall SFE is typically the average over the whole star-forming GMC, perhaps also including regions where no star formation occurs and thus with significantly low local efficiency. Indeed, there is an increase in SFE with average cloud density, from several percent in whole giant molecular clouds to $\sim$\,30\% or more in dense cluster-forming cores. This difference can be  understood as the result of the hierarchical cloud structure \citep{Elmegreen2008}. {The average SFE in the GMCs of the star-forming complexes, which are responsible for 33\,percent of the free-free emission in the Milky Way, was found to be significantly higher than the Galaxy-wide average \citep{Murray2011}.} SFE is also found to depend on the metallicity of the gas, based on the dependence of stellar winds on metallicity. Lower metallicity winds inject less momentum and kinetic energy into the ambient gas, and thus, are less efficient in expelling the gas from the clump, leading to higher efficiency \citep{Dib2011}. SFE is also expected to depend on the proto-cluster clump mass, for a given star formation history in the clump and a fixed feedback coupling efficiency with the ambient gas \citep{Dib2013}. 

Numerical work suggests that supersonic turbulent flows controls star formation with inefficient, isolated star formation being the signpost of turbulent support, while efficient, clustered star formation occurring in its absence \citep{MacLowKlessen2004}. The density contrast in turbulent isothermal gas is also found to scale positively with the turbulence Mach number \citep[e.g.,][]{Konstandin2016}. Turbulent systems are typically  gravitationally unstable \citep[][]{McKeeOstriker2007ARA&A, Hopkins2013}, with supersonic turbulence driving  their density distribution \citep[e.g.,][]{Klessen2011EAS, Hopkins2012}. Bound clusters appear to form in the densest parts of a self-gravitating cloud complex that is structured by turbulence \citep{Elmegreen2014LEGUS}. The reason is that high-density peaks in a GMC usually have free-fall times short enough to achieve the necessary high SFE for the formation of bound clusters \citep{Kruijssenetal2012}. On the other hand, low-density gaseous condensations, which are more common in galaxies, drive dispersed star formation that is spatially clustered yet unbound. This has led to the suggestion that loose stellar systems may represent the most common form of clustered star formation, as has been predicted by theory \citep{Kruijssen2012}, and shown by significantly large numbers of dispersed young populations related to lower gas density and SFE within the same cloud \citep[][]{GouliermisHony2017}.

Various studies have suggested that pressure is the dominant factor in determining the kind of unit that is formed, e.g., bound super cluster, open cluster, or association, \citep[e.g.,][]{ElmegreenEfremov1997, AshmanZepf2001}. Regions of low ISM pressure are expected to form relatively fewer bound clusters and more unbound associations \citep[e.g.,][]{Elmegreen2008, Meidt2013PAWS}.  The difference between star formation in bound clusters and star formation in loose groupings is attributed to a difference in cloud pressure, with higher pressures forming more tightly bound clusters (still equally massive as loose associations).  This is observed in M\,83, where cluster formation is found to become less efficient, as gas pressure radially declines towards the outskirts of the galaxy \citep{Adamo2015}. {In the inner disk of the Milky Way \cite{Ragan2016} found that the prevalence of star formation in  clouds revealed by the {\sl Herschel} Hi-GAL survey \citep{Molinari2010} declines with Galactocentric radius. These authors suggest that physical properties of the Galaxy, such as metallicity, radiation field, pressure or shear, may affect the 
GMCs SFE.}  Gas pressure may also explain why super star clusters are found in starbursts and merging galaxies but not in galaxies like the Milky Way, where star formation is driven by such internal processes \citep{PortegiesZwart2010ARAA}.




\begin{figure*}[t]
\centering
\includegraphics[width=0.75\textwidth]{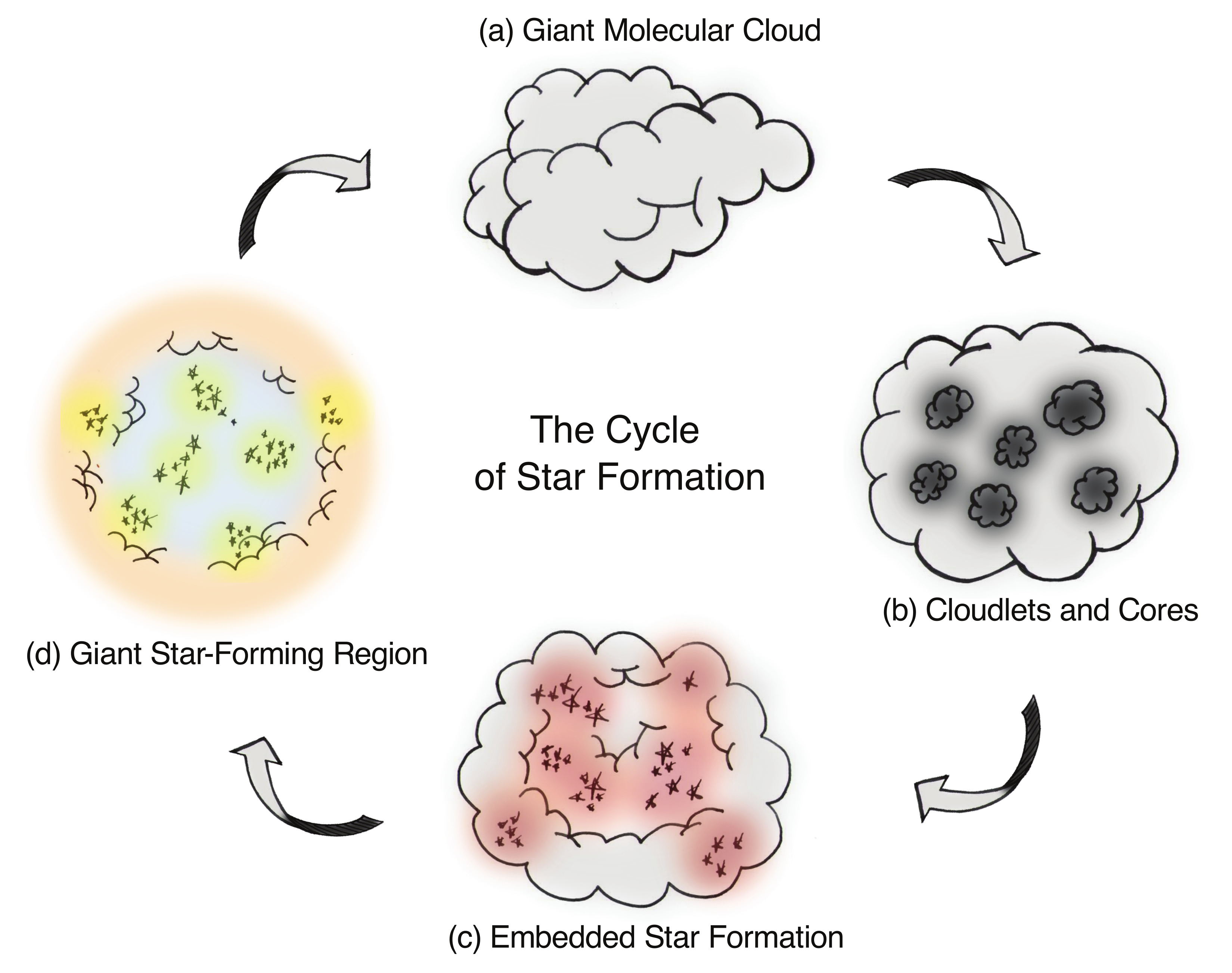} 
\caption{\small {\em A cartoon of the cycle of the star formation process}. (a) A molecular cloud under the right environment conditions (density, temperature, pressure) is locally compressed to initiate star formation. (b) Individual ``centers'' (cores and cloudlets) with density higher than the local average will form. Self-gravity will lead to the formation of stars and clusters inside these centers, which are not necessarily bound to each other. (c) The cloud becomes sub-structured into individual proto-clusters (compact or not) still embedded in molecular gas. Stellar radiation heats the circumcluster interstellar gas and dust, which reemit in the infrared. (d) Radiative and mechanic feedback from massive satrs and super-novas eradicates much of the remaining, less-dense molecular material, while dense ``inactive'' parts of the cloud are still present. A giant star-forming region (or star-forming complex) is formed. Typical demonstration of stellar feedback is the formation of bubbles and photodissociation regions (PDRs) at the locations where feedback meets the cloud. The formed sub-clusters may dissolve after gas expulsion while their cores contract, feeding the inter-cluster region with young populations, or merge into forming a young massive cluster. This complex interplay between gas and young stars dumps metals and energy into the interstellar matter to be eventually used in subsequent star formation events. \label{f:sfcycle}}
\end{figure*}





\subsection{The formation of Unbound Systems\label{originOB}}

 \cite{Ambartsumian1955} opens the discussion about the origin of stellar associations and argues that poor clusters of O- and B-type stars may dissolve by the time their stars are old, probably turning into OB associations in $10^6$ years, although, as he notes, {\sl not all associations have necessarily arisen from clusters}. Early observations of stars forming at the boundaries of Galactic molecular clouds
and HII regions \citep{Habing1972, Ladaetal1976} motivated the classic theory regarding the formation of OB associations.
This theory proposes that OB associations form through triggering, prompted by the ionized regions produced by previous generations of OB stars. Star formation is thus self-propagating, with one
generation of stars triggering the formation of the next {\citep{MuellerArnett1976, ElmegreenLada1977, GerolaSeiden1978}}. In this scenario the shocked layer where the new generation of OB stars forms is moving away from the old generation of stars, with velocities of a few km\,s$^{-1}$. The new OB-group will be thus unbound from its parental group, in agreement to the observed dynamics of Galactic OB associations. The self-propagating star formation scenario favors the formation of high-mass stars, while low-mass stars would form
spontaneously in the rest of the cloud, suggesting a time delay between the formation of low- and high-mass stars, and in locations which would be physically well-separated. This prediction differs from the observed stellar content of OB associations \citep{Garmany1994, Brown1999ASIC, Massey2003ARA&A}, which show a fully populated stellar IMF within their sub-clusters  \citep{deGeus1992, PreibischZinnecker1999, Briceno2007ppv}. Nevertheless,  observations of HII bubbles do suggest a causal relation between the first generation of stars located {\sl in} the bubble and the second generation of embedded clusters located {\sl at its rim} \citep{DeharvengZavagno2011}.

Other theoretical scenarios suggest a closer relation between the formation of unbound associations and standard clustered star formation. For instance, \cite{ClarkBonnellZinnecker2005} simulate the formation of OB associations 
as a series of clusters forming in an {\sl unbound} GMC\footnote{The supporting kinetic energy
of the GMC in these simulations is considered to be twice as large as the cloud's 
self-gravity.}. The cloud after forming the clusters, which are unbound from one another\footnote{Sub-clusters are unbound from each other, due to the fact that the flows that form them are also unbound from each other.}, disperses  within roughly two crossing times, supporting the idea that star formation is a rapid process. In contrast to previous simulations \citep[e.g.,][]{BonnellVB2003}, the formed clusters, which
behave as OB subgroups, expand away from one another and from their mutual center of mass, rather than merging into a larger cluster. This star formation scenario proposes that unbound clouds can form OB associations through the clustered star formation mechanism. OB associations may be thus the result of the formation of many localized bound regions in an overall unbound cloud. A naive cartoonist's depiction of such a process is shown in Fig.\,\ref{f:sfcycle}. Large tidal effects may disrupt then the unbound cloud. Support to this scenario is provided by a study of the low-mass YSOs surface densities in star-forming regions in the Solar neighborhood, which found no evidence for discrete modes of star formation, such as, e.g., clustered versus distributed \citep{Bressert2010}. This investigation showed that the fraction of stars formed in clusters is very sensitive to the adopted cluster definitions, and that only a small fraction of less than 26\,per\,cent of stars are formed in environments dense enough for their formation and evolution to be affected by their close neighbors. Its  results suggest that both bound and unbound systems may have common origin. 


Star formation in a turbulence-supported GMC may also  lead to the formation of a 
massive compact cluster through merging of individual sub-clusters. In simulations of a bound 10$^3$\,M$_\odot$ GMC, which was initially supported against collapse by a turbulent velocity field,
the dissipation of the large-scale supersonic flows produces distinct sub-clusters, each of them 
including a central massive star \citep{BonnellVB2003, BonnellBV2004}. Mass segregation and the observed field stellar IMF were reproduced in these sub-clusters through competitive accretion. In these simulations the initially bound
cloud leads to the formation of sub-clusters which are themselves bound to one another, after 
the dissipation of the turbulent energy. These clusters merge within roughly twice the free-fall time for the original cloud, not being able to form an OB association, but instead one large cluster. The comparison of this scenario for clustered star formation to that discussed in the previous paragraph indicates that whether the outcome of star formation will be a bound or unbound stellar concentration seems to depend on the initial conditions.  

The conditions required for the formation of bound or unbound star clusters are explored by \cite{GeyerBurkert2001}, who found that whether the newly-formed star cluster or pieces of it remain gravitationally bound or not depends mainly on the local SFE and not on the efficiency averaged over the whole cloud. {However, \cite{Smith2011MNRAS414} among others, argue that SFE is not the defining factor for cluster survival. There is, thus, no consensus in the literature about what determines cluster survival.} The importance of the initial conditions to the fate of the newly-formed sub-clusters is also investigated in a series of numerical simulations of star-forming regions by \cite{Parker2014}. The synthetic stellar distributions were constructed with a spatial scale of few pc to be initially fractal (in order to introduce sub-structure) and with various initial global virial ratios (kinetic over total potential energy of the stars). Sub-virial and virialized regions are found to evolve into mass segregated bound clusters, while super-virial distributions usually evolve to form unbound associations. In all regions massive stars attain local densities, which are higher than the median. Dense bound regions separate from dense  unbound ones within a timescale of $\sim$\,5\,Myr. The former evolve within length-scales of few pc, while the latter expand to areas extending up to few 10 pc scales. More recent simulations suggest that the global virial parameter of the star-forming region is the key quantity in determining the formation and subsequent evolution of bound clusters \citep[e.g.,][]{LeeGoodwin2016}. Star-forming regions should be, thus,  investigated as a whole, rather than on individual substructures basis \citep[see, e.g.,][]{Gouliermis2014}. 

The situation is more complex when one includes the dynamics of molecular cloud formation. According to the {\sl bathtub} model the question of whether a cluster forms as a bound or unbound entity depends on the timescale on which gas accretes into the molecular cloud, compared to the star formation timescale \citep[][]{Burkert2017}. The model predicts that the star formation rate of a molecular cloud is not determined by its mass or internal collapse timescale, but by the accretion rate of new gas. The importance of initial conditions to the statistical properties of the produced proto-clusters is investigated by \cite{Girichidis2012MNRASIII}, who found that the clusters vary significantly in size, mass and number of protostars, and show different degrees of substructure and mass segregation. Magnetic fields are also important. Stronger magnetic fields are found to reduce the rate of dense core formation, and therefore the formed clusters tend to be more sparse and unbound \citep{PriceBate2008, Dib2010, BasuDapp2010}. 

The formation of the most massive stars at the centers of compact clusters, coined with the term {\sl primordial mass segregation}, seems to be a common outcome of star formation \citep[][]{BonnellDavies1998, deGrijs2002, Gouliermis2004}. In simulations of star-forming molecular clouds massive stars are only formed at the bottom of bound potential wells, where large-scale collapse could channel large amounts of mass onto a protostar \citep[e.g.,][]{Smith2009}, suggesting that massive stars could only form in the bound regions of molecular clouds. This is in agreement with observations, where massive stars are found systematically at the centers of compact systems of various masses, even under conditions of extreme isolation \cite[i.e., unrelated to any extended star-forming region, see, e.g.,][]{Stephens2017}.  Feedback in the form of ionization and winds from these stars influences the dynamics, structure and SFE of their natal clouds. 
Photoionization from O-type stars in turbulent star-forming clouds affects the structure of the gas, creating
large and clear bubble structures and pillars,  reduces the star formation efficiency  and  rate, and triggers local star formation \citep{Dale2013V, Dale2013III}. {Momentum-driven OB-star stellar winds, on the other hand,  are found to be considerably less effective in disrupting their host clouds than the ionizing radiation, originated in HII regions \citep{Dale2013IV}.}

The stellar IMF, the mass distribution of stars
at the time of their formation is a cornerstone of our understanding of star formation. Its shape and dynamic range 
and the variability of these parameters play essential role in determining the physics of star formation. 
{The shape of the stellar field IMF appears to be approximately invariant across various environments \citep[][]{Kroupa2002Sci, Chabrier2003PASP, Bastian2010ARA&A}.  Generally, the balance of evidence is in favor of a \cite{Salpeter1955} power-law at high masses that flattens at the so-called ``knee'' of the IMF, at around 1~M{\solar}. However, the stellar field IMF is subjected to a range of dynamical 
and environmental effects within the clusters (e.g., binary disruption, proximity to radiation fields from OB stars, mass segregation), and concerns stars that have been ejected from a variety of cluster environments. As a consequence this IMF is the average result of various processes, and possibly the outcome of mixing of environments over the lifetime of the Galactic disk. This may explain the apparent approximate invariance of the stellar field IMF, with the most possible causes of variations being averaged out in both location and time.
In contrast, the IMF deduced from an individual star cluster is based on a snapshot of the evolution of the cluster up to that time. The stellar 
IMF of a young cluster will, thus, depend on its dynamical state, the fraction of binaries that 
formed and have been disrupted, the influence of massive stars, as well as other environmental 
effects. Indeed, different classes of simulations \citep[e.g.,][]{Bonnell2006MNRAS, PadoanNordlund2007ApJ} locate the IMF knee at a mass related to the thermal Jeans mass of the parental molecular cloud, and various studies predict different  characteristic mass scale in the IMF \citep[e.g.,][]{Larson2005MNRAS, Elmegreen2008ApJ}. These findings suggest that the stellar IMF may be variable, and subject to environmental conditions,  contradicting the ``universality'' of the field IMF. Indeed, significant cluster-cluster variations of the IMF are recently found among the population of young clusters in the Milky Way \citep{Dib2017a}.}

\greybox{The formation of UYSS is tightly related to clustered star formation within an overly unbound molecular cloud. Supersonic turbulence introduces fluctuations in the density distribution of GMCs, with localized high density regions producing bound clusters with high SFE, while low-density regions drive the formation of dispersed stellar populations that are spatially clustered yet unbound. Turbulence, along with self-gravity, are the main drivers of the observed structure and its hierarchical pattern in both the gas and young stars.  Star formation across fractal gaseous structures highlights the turbulence cascade in the gas by producing stars wherever the gas is dense enough for self-gravity to dominate.}

The observed high-mass truncation of the IMF, i.e., the maximum 
stellar mass, in a star cluster is suggested to scale with the cluster mass  \citep{OeyClarke2005ApJ}. This relation may 
be explained by the hypothesis that clusters build their IMFs with {stochastic sampling} of a universal stellar 
IMF \citep[][see also \citealt{Elmegreen2006ApJ}]{ParkerGoodwin2007}, or by the contrary hypothesis of {optimal IMF sampling},
which assumes that stars derive their mass directly from the mass reservoir of the parent cluster  \citep{WeidnerKroupa2006MNRAS}. 
Recent results on isolated compact clusters around massive YSOs provide evidence in favor of the optimal sampling scenario, but they concern 
the extreme case of clustered star formation in isolated environments \citep[][]{Gouliermis2018MemSAIt}. It is still unexplored if the IMF depends on how bound a young cluster is. For example, there are theoretical indications that unbound regions may not form fully-populated IMFs \citep{Bonnell2011}. Also variations are reported in the IMF within the same star-forming region, where differences  appear between the extended association and its dominant compact cluster. As a consequence, caution is advised in the definition of the boundaries of individual systems for the construction of their stellar IMF  \citep[see, e.g.,][]{Gouliermis2002}. This also applies to the determination of the structural characteristics of stellar systems, since the result is extremely sensitive to the considered stellar populations \citep[e.g.,][]{Bressert2010, Schmeja2011AN}.


\section{Concluding Remarks \label{s:conclusion}}

Stars form in groups that demonstrate a diversity in length-scales and degrees of gravitational self-binding  \citep[][]{Kontizas1999IAU}. The distinction between 
bound and unbound stellar systems is made upon the stability of stellar systems in terms of two factors: (1) Their structural characteristics, such as their volume density and radiative pressure, and (2) their surrounding environments, such as gas pressure and disruption by close-by clouds. From the original analyses of \cite{Bok1934} and \cite{Spitzer1958} to those of \cite{Blaauw1964} and \cite{LadaLada1991}, typical limiting values of volume density for the stability of stellar systems are set to $\leq\,0.1$\,M$_{\odot}$\,pc$^{-3}$ for unbound systems and $\geq\,1.0$\,M$_{\odot}$\,pc$^{-3}$ for bound clusters. A large variety of newly-born stellar systems span across the whole extent between and beyond these density limits, from low-mass compact star clusters and young massive clusters to elongated unbound small systems and large loose stellar complexes. These stellar structures host the young stars in a galaxy, and feed its general field as they evolve. They are the fundamental building blocks of galaxies, being a major source of light (photon radiation), and metals (via stellar winds and supernova explosions) in the universe. Independently of their morphology, evolutionary stage, and structural behavior, unbound young stellar systems (UYSS) host important information about the star formation process and the resulted stellar populations. They are thus the rosetta stone in deciphering the recent star formation history, chemical enrichment, and dynamical evolution of their parental galaxy. This review has focused on our knowledge about UYSS, and particular on their characteristics, their formation, and their relation to star clusters, as their compact parts.

Numerous surveys of Galactic star-forming regions  \citep[see, e.g.,][]{Reipurth2008hsf1, Reipurth2008hsf2} indicate GMCs 
as the basic unit where stars form. Undoubtedly the formation of stellar systems within GMCs is intimately connected to the star formation process
\citep[e.g.,][]{Lada2010RSPTA.368}, which is governed by the complex interplay between gravity, supersonic turbulence, 
thermal pressure, and magnetic fields, all ``stored'' in the interstellar matter \citep[e.g.,][]{Pudritz02}. 
Turbulence is identified as the main physical process that regulates clustered star formation \citep[e.g.,][]{MacLowKlessen2004, McKeeOstriker2007ARA&A}.
According to theory stellar systems build up hierarchically, in the sense of {\em systems forming within systems} \citep[e.g.,][]{KlessenBurkert2000, Clarke2000ppiv, BonnellVB2003}. This {\em hierarchy} seems to be naturally inherited 
from the hierarchical structure of the ISM, the most likely source of which is the combination of mechanisms such as agglomeration with fragmentation 
\citep{McLaughlinPudritz1996ApJ}, self-gravity \citep{deVega1996Natur}, and turbulence \citep[][]{ElmegreenScalo2004}. Star formation appears to proceed hierarchically in both space and time. The duration of star formation tends to increase with the size of the region as the turbulet crossing time, suggesting that small compact systems form stars faster than larger extended structures \citep{EfremovElmegreen1998}. On the other hand, early cluster dissolution, which occurs through the disruption of the surrounding gaseous environments by the newly-formed stars, may lead not only to the rapid disruption of a cluster-forming core, but also to the dispersal of the star-forming GMC \citep[e.g.,][]{Whitworth1979}, influencing the dynamics of the region. It, thus, stands to reason that both factors, i.e., 1) the complex star formation process and 2) the subsequent environmental and dynamical conditions, apparently produce the observed variety of UYSS in size, shape, and compactness. 

On small, sub-GMC scales, where gas density is higher than the GMC average \citep[e.g.,][]{PadoanNordlund2011},
star formation is a strongly clustered and very efficient process \citep{Hopkins2013MNRAS}. On the other hand, regions of low gas density in GMCs supported by turbulence produce stars in an inefficient, isolated manner \citep{MacLowKlessen2004}, leading to the formation of dispersed stellar distributions, which are spatially clustered yet unbound. The formation of UYSS can be, thus, explained as the formation of several localized bound sub-clusters in an overall unbound cloud \citep[e.g.,][]{ClarkBonnellZinnecker2005, Kruijssen2012}. Observations indicate that the typical mass fraction of dense gas is rather small, about 10\,per\,cent of the total molecular cloud mass \citep[see, e.g.,][and references therein]{Vutisalchavakul2016}, while low-density gaseous structures appear to be more common in galaxies. The formation of unbound systems may, thus, be a very common form of clustered star formation. Star formation may also proceed within the GMC in an unbound way through self-propagation, with one generation of stars being induced by another \citep{ElmegreenLada1977}. Independently of the formation mechanism, the dynamical evolution and interactions between the bound sub-clusters formed along with the dispersed populations, i.e., their disruption and/or merging, as well as feedback from the massive stars, will affect the subsequent dynamics, structure and overall SFE of the region \citep[e.g,][]{BonnellVB2003, Dale2013V}. The formation of extended unbound stellar agglomerations or compact massive young clusters as the outcome of this process seems to depend on the initial conditions, such as the density distribution and velocity structure of the natal GMC \citep[e.g.,][]{Girichidis2012MNRASIII, Parker2014}. 

There are several issues related to the formation and evolution of compact stellar clusters, 
such as the cluster mass function, the fraction of star formation happening 
in bound clusters ($\Gamma$) and  the maximum cluster mass, to mention few. 
While these issues may be related to hierarchical star formation, they are not entirely related to the scope of this paper and therefore not sufficiently addressed. For a detailed account 
of these topics the reader can refer to the review articles compiled in the recent monograph {\em The Birth of Star Clusters} \citep{Stahler2018ASSL}. 
Here is worth mentioning that the power-law shape of extragalactic cluster mass functions is interpreted as the product of hierarchy in both space and time
\citep[e.g.,][]{Messa2018LEGUS1}, in agreement with the idea that compact clusters are (the densest) parts of larger hierarchical stellar structures. These mass functions do not seem to follow a power-law across their entire dynamic range, but there is a confirmed dearth of clusters at high masses \citep[e.g.,][]{Adamo2017LEGUS}. Indications that the mass limit of the cluster mass function truncation may depend on the galactic environment \citep[][]{Johnson2017PHAT} suggest a clear relation between the formation of clusters (as the high density peaks of the hierarchy) and the global star formation properties of their host galaxies.

Internal processes in the GMCs that lead to the formation of stellar complexes are most likely related to their galactic environments and the dynamics of their hosting galaxies. For instance, gravitational instabilities in the stars, and transient instabilities in the gas that trigger star formation and generate turbulence usually result to long irregular or flocculent spiral arms \citep{Elmegreen2011SFSpiralArms}. Global spirals force the gas into a dense molecular phase in dust lanes and organize it in a spiral pattern. Therefore, processes that lead gas to form stars shape the observed galactic spiral structure \citep{Dobbs2014PASA}. Recent surveys of large-scale UYSS in nearby galaxies reveal this structure and indicate that self-similar hierarchy in stellar clustering extends from few pc scales up to those of galactic disks \citep[e.g.,][]{Gouliermis2017LEGUS}. These studies suggest that UYSS are possibly the fundamental link between GMC-wide and galaxy-scale star formation. However, the systematic investigation of star formation on multiple scales across whole star-forming galaxies is a relatively unexplored topic. 

Moreover, there are few issues that are not well understood yet. How important is self-gravity (and collapse) compared to turbulence? 
Can self-gravity alone (with no pressure) produce the observed hierarchical clustering? Is turbulence indeed the dominant mechanism with gravity playing a role only at the very densest scales? What would be the stellar clustering pattern for each of these extreme cases? What is the connection between the empirical relation between SFR and the gas density, the Kennicutt-Schmidt law \citep{Schmidt1959, Kennicutt1998}, to hierarchical star formation? Do hydrodynamical simulations of star formation predict the observed structure of stars without resolving all its scales?  These are few of the questions that require investigation. 
Eventually, linking the ``microscopic'' processes of the formation of stellar systems (turbulence, gravity, feedback) to the ``macroscopic'' parameters of star-forming galaxies (star formation rate, gas reservoir, dynamics), in order to establish the relation between stellar clustering statistics and galaxy-wide properties, is an exciting field that hopefully will give us a better understanding about how stars form in groups.

\section*{Ackowledgements}

I would like to thank PASP Editor-in-Chief, Jeff Mangum, for his help and patience and the unknown referee for her/his detailed and helpful review of this article. I would also like to thank 
Javier Ballesteros-Paredes, Ian Bonnell, Andi Burkert, Paul Clark, Sami Dib, Clare Dobbs, Diederik Kruijssen, Rowan Smith, and Nick Wright for sharing their insightful thoughts about
unbound star formation. Sincere thanks go to Bruce Elmegreen and Ralf Klessen for reading the manuscript and providing critical comments for its improvement. 
Financial support by the German Research Foundation (DFG) through program GO\,1659/3 {\sl ``Clustered Star Formation on Various Scales: An extensive panchromatic stellar survey of star-forming galaxies in the Local Group with the Hubble Space Telescope''}, and the German Aerospace Center (DLR) and the Federal Ministry for Economic Affairs and Energy (BMWi) through program 50OR1801 {\sl ``MYSST: Mapping Young Stars in Space \& Time''} is kindly acknowledged. Most importantly, I thank my beloved wife and children for their continuous support of my research work in spite of all the work-related uncertainties.

\renewcommand\refname{\large\fontfamily{phv}\selectfont{Bibliography}\vspace*{-.75truecm}}

\end{document}